\documentclass{aa}
\usepackage{graphicx,natbib}
\usepackage[latin1]{inputenc}
\usepackage[T1]{fontenc}
\usepackage{txfonts}
\usepackage{changebar}
\usepackage{bm}
\usepackage{units}

\defcitealias{2005A&A...434..343A}{A05}
\defcitealias{2013A&A...549A..44F}{F13}
\defcitealias{2009A&A...501.1139M}{M09a}
\defcitealias{2009A&A...501.1161M}{M09b}
\defcitealias{2012A&A...547A.111M}{M12b}
\defcitealias{2012A&A...547A.112M}{M12a}

\def\mearth{M_{\oplus}}

\def\tdisk{T_{\rm disc}}

\def\f1{f_{\rm I}}
\def\msun{M_\odot}
\def\mstar{M_{\rm star}}

\def\re{R_{\mathrm {E}}}

\def\beq{\begin{equation}}
\def\eeq{\end{equation}}

\def\rhopla{\rho_{\rm planetesimals}}
\def\rpla{R_{\rm planetesimals}}

\def\rhogas{\rho_{\rm gas}}

\def\vrel{v_{\rm rel}}
\def\re{Re}

\def\mdisk{M_{\rm disc}}
\def\ac{a_{\rm C}}
\def\rin{r_{\rm inner}}
\def\rzero{r_0}
\def\rcap{R_{\rm cap}}

\def\tdisk{T_{\rm disc}}

\def\hdisk{H_{\rm disc}}

\def\mearth{M_\oplus}
\def\msun{M_\odot}

\def\rhill{R_{\rm Hill}}
\def\rhills{R_{\rm Hill}}

\def\aplanet{a_{\rm planet}}
\def\mplanet{M_{\rm planet}}

\def\mcore{M_{\rm core}}

\def\mearth{M_\oplus}

\def\mcore{M_{\rm core}}

\def\aplanet{a_{\rm planet}}

\def\rhill{R_{\rm H}}

\def\simgr{\,\hbox{\hbox{$ > $}\kern -0.8em \lower 1.0ex\hbox{$\sim$}}\,}
\def\simle{\,\hbox{\hbox{$ < $}\kern -0.8em \lower 1.0ex\hbox{$\sim$}}\,}
\def\beq{\begin{equation}}
\def\eeq{\end{equation}}

\def\vect#1{\overrightarrow{#1}}
\renewcommand{\vect}[1]{\ensuremath{\mbox{\boldmath$#1$}}}

\def\simgr{\,\hbox{\hbox{$ > $}\kern -0.8em \lower 1.0ex\hbox{$\sim$}}\,}
\def\simle{\,\hbox{\hbox{$ < $}\kern -0.8em \lower 1.0ex\hbox{$\sim$}}\,}
\def\beq{\begin{equation}}
\def\eeq{\end{equation}}

\def\apjl{ApJ}                
\def\apjs{ApJS}               
\def\ao{Appl.Optics}          
\def\aap{A\&A}                

\def\({\left(}
\def\){\right)}
\def\<{\left<}
\def\>{\right>}

\def\({\left(} 
\def\){\right)} 
\def\<{\left<} 
\def\>{\right>} 

\def\bc{\begin{changebar}}
\def\bce{\begin{center}}
\def\beq{\begin{equation}} 

\def\bi{\begin{itemize}}

\def\btab{\begin{tabular}{p{1.7cm}p{12cm}p{1.5cm}}}
\def\bt2{\begin{tabular}{p{1 cm}p{4.5cm}p{10cm}}}

\def\ec{\end{changebar}}
\def\ece{\end{center}}
\def\eeq{\end{equation}} 
\def\ei{\end{itemize}}

\def\etab{\end{tabular}\\}

\def\mH2{m_\mathrm{H_2}}

\def\dmax0{\rho_\mathrm{max}}
\def\dmaxS0{\Sigma_\mathrm{max}}

\def\rH2{r_\mathrm{H_2}}

\def\r0max{r_\mathrm{0max}}

\def\s0{\sigma_\mathrm{0}}

\def\xp0{x_{\rm{M0}}}

\def\z0max{z_\mathrm{0max}}

\def\apjl{ApJ}                
\def\apjs{ApJS}               
\def\ao{Appl.Optics}          
\def\aap{{\it A\&A}}                

\begin{document}

\title{Theoretical models of planetary system formation: mass {\it vs} semi-major axis}

\author{Y. Alibert \inst{1,2}
       \and
       F. Carron \inst{1}
       \and
       A. Fortier \inst{1}
       \and
       S. Pfyffer \inst{1}
       \and
       W. Benz \inst{1}
       \and
       C. Mordasini \inst{3}
       \and
       D. Swoboda \inst{1}
     }
\offprints{Y. Alibert}
\institute{Physikalisches Institut  \& Center for Space and Habitability, Universit\"at Bern, CH-3012 Bern, Switzerland, \\
        \email{Yann.Alibert@space.unibe.ch, Frederic.Carron@space.unibe.ch, Andrea.Fortier@space.unibe.ch, Samuel.Pfyffer@space.unibe.ch, WBenz@space.unibe.ch, David.Swoboda@space.unibe.ch}
           \and
        Observatoire de Besan\c con, 41 avenue de l'Observatoire, 25000 Besan\c con, France\\
        \email{alibert@obs-besancon.fr}
      \and
      Max-Planck Institute for Astronomy, K\" onigstuhl 17, 69117 Heidelberg, Germany
    \email{mordasini@mpia.de}}

\date{Received 12 April 2013 / Accepted 08 july 2013}

\abstract
{Planet formation models have been developed during the last years in order to try to reproduce the observations of both the solar system, and the extrasolar planets. Some of these models have partially succeeded,
focussing however on massive planets, and for the sake of simplicity excluding planets belonging to planetary systems. However, more and more planets are now found in planetary systems. 
This tendency, which is a result of both radial velocity, transit and direct imaging surveys, seems to be even more pronounced for low mass planets.
 These new observations require the improvement of planet formation models, including new physics, and considering the formation of systems.}
{ In a recent series of papers, we have presented some improvements in the physics of our models, focussing in particular on the internal structure
of forming planets, and on the computation of
the excitation state of planetesimals, and their resulting accretion rate. In this paper, we focus on the concurrent effect of the formation of more than one planet in the
same protoplanetary disc, and show the effect, in terms of global architecture and composition of this multiplicity.}
{We use a \textit{N}-body calculation including collision detection to compute the orbital evolution of a planetary system. Moreover, we describe
the effect of competition for accretion of gas and solids, as well as the effect of gravitational interactions between planets.}
{We show that the masses and semi-major axis of planets are modified by both the effect of competition and gravitational interactions. We also present the effect of the assumed number
of forming planets in the same system (a free parameter of the model), as well as the effect of the inclination and eccentricity damping. { We find that the fraction of ejected planets
increases from nearly 0 to 8 \% as we change the number of embryos we seed the system with from 2 to 20 planetary embryos. Moreover, our calculations show that, when considering planets more massive than
$\sim 5 \mearth$, simulations with 10 or 20 planetary embryos give statistically the same results in term of mass function and period distribution.}}
{}
\titlerunning{Theoretical models of planetary system formation}
\authorrunning{Y. Alibert et al.}

\maketitle

\section{Introduction}
\label{sec:introduction}

Since the pioneering discovery of 51 Peg b~\citep{1995Natur.378..355M}, the first extrasolar planet orbiting a solar type star, the statistic
of planetary observations has shown an exponential growth. This tendency has been amplified during the last years by the growing number
of planetary candidates discovered by Kepler~\citep{2011ApJ...728..117B}. One interesting feature in these planetary observations, which is a characteristic
of both radial velocity and transit surveys, is that the number of planetary systems is also growing extremely rapidly~\citep[see e.g.][]{2011A&A...528A.112L,2011arXiv1109.2497M,2011Natur.470...53L}. Such planetary systems are very interesting for planet formation theory, since they can provide constraints on the processes acting
during planet formation. For example, the presence of resonant systems seems to be very probably linked to migration during
planet formation. On the other hand, from the theoretical point of view, and as we shall see in this paper, the formation of a planetary system is a problem 
significantly more difficult to solve than the formation of an isolated planet.

Indeed, the formation of a planetary system involves not only the formation process of the individual planets themselves,
a process which is far from being fully understood for the moment, but also all the interactions between these planets.
Among all these interactions, one can mention:
\begin{itemize}
\item Growing planets, in particular when they are close to each other, are competitors for the accretion of solids and gas.
\item Planetesimal accretion can be strongly perturbed by the presence of neighboring planets that can generate density waves of solids due to
the excitation they produce in the random velocity of planetesimals~\citep{2010A&A...521A..50G,2011A&A...532A.142G}. Depending on the location of planets, the planet mass ratio,
the density profile of the disc, and the size of planetesimals, the accretion rate of planetesimals can be reduced or enhanced.
\item The formation of a planet in the wake of another one is strongly perturbed. This was shown, using simplified models, in the case of the Solar System~\citep{2005ApJ...626L..57A} and the HD~69830~system~\citep{2006A&A...455L..25A}.
\item Gravitational interactions between forming planets modify their migration, and may lead to mean motion resonances systems.
\item Collisions between protoplanets, or ejections of some planets,  are likely to occur during the whole formation phase. 
\item Gap formation in a disc is modified when more than one planet is present. Moreover, the merging of two neighboring gaps leads to 
modification of migration~\citep[see][]{2001MNRAS.320L..55M}.
\item When a planet forms a gap, the {outer boundary of the gap} may represent a place of very reduced type~I migration, thus acting as a planet trap~\citep{2006ApJ...642..478M}.
 \end{itemize}
 
 In a forthcoming series of papers, we intend to improve our planet formation models,  in terms of both the physical and numerical treatment of some of the important
 processes involved in the formation of a system (migration, protoplanetary disc structure and evolution, and planet internal structure), and  to include some 
 of the afore-mentioned interaction effects. Improved disc models and planet internal structure models have been described in \citet[\citetalias{2012A&A...547A.112M} and \citetalias{2012A&A...547A.111M} in the following]{2012A&A...547A.112M,2012A&A...547A.111M} and \citet[\citetalias{2013A&A...549A..44F} in the following]{2013A&A...549A..44F}, improved migration models are presented in Dittkrist et al. (in prep). In this paper, 
 we focus on the effect of forming more than one planet in the same protoplanetary disc. We present the numerical approach used to compute the gravitational
 interactions and collisions between planets,  the treatment of the competition for solid accretion,
 and  the differential effect of having more than one planet growing and migrating in the same disc. 
The interactions between forming planets, mediated through the gas component of the protoplanetary disc (a first planet modifying the protoplanetary disc - e.g. by gap formation or spiral wave generation - this leading to e.g. a modified migration of a second planet), 
 will be considered in another work, as it requires the development of new numerical models. As a consequence, although
 we consider populations of planets that are not fundamentally different from observed populations, our results should be
  considered only as a step toward a global understanding of planetary population, and will likely be improved and modified in the
 future. The specific application of our models to the case of the Solar System will be considered in a future paper, since the process
 of gap merging is likely to have played an important role in this case~\citep[see][]{2011Natur.475..206W} .
 
 The paper is organized as follows: we present in Sect. \ref{formationmodel} a summary of the most important physical features of our models, summarizing the
 work presented elsewhere \citepalias[for details see][]{2012A&A...547A.112M,2012A&A...547A.111M,2013A&A...549A..44F}. This section is presented for the sake of completeness, and can be skipped by readers having read the papers mentioned above.
 In Sect. \ref{nbody}, we describe the computation of the planet's orbital evolution (including planet-planet gravitational interactions  and disc planet interactions) and collision detection.
 In Sect. \ref{competition}, we present our treatment of the competition for the accretion of solids and gas. In Sect.  \ref{results} we present the results,
 considering both an example of 10-planet system formation models, and the results of planetary population synthesis, 
 comparing the case where only one planet forms in the protoplanetary disc, and the case where multiple planets form. As we shall see later, the number of
 planets growing in the disc is a free parameter of the model, and we will present in this paper the case where 1, 2, 5, 10 and 20 planetary embryos grow and migrate in the disc, and
 will discuss the sensitivity of the results to this number. Finally, in Sect. \ref{discussion}, we discuss our results and limitations and future developments of the models.

\section{Formation model}
\label{formationmodel}

The formation model consists of different modules, each of them computing one important class of physical processes involved during the
formation of a planetary system. These modules are related to the protoplanetary disc structure and evolution (including both vertical and radial structure),
the computation of the planetesimals' dynamical properties and accretion rate, the planets' internal structure, and the dynamical interactions between planets and between the disc and the planets. These different modules have been already described elsewhere \citep[e.g.][hereafter~\citetalias{2005A&A...434..343A,2012A&A...547A.112M,2012A&A...547A.111M,2013A&A...549A..44F}]{2005A&A...434..343A}.

\subsection{Protoplanetary disc: gas phase}
\label{disc}

\subsubsection{Vertical structure}

The structure of protoplanetary discs is complex, and different effects may be important.
There could be irradiation, and also the presence of a dead zone. In our model, the vertical disc structure is
computed by solving the equations for hydrostatic equilibrium, energy conservation, and 
diffusion for the radiative flux~\citepalias[see][]{2005A&A...434..343A,2013A&A...549A..44F}.

This calculation provides us with the vertically averaged viscosity as a function of the surface density in the disc.
In the models presented here, we assume that the local viscosity is given by the Shakura-Sunyaev approximation~\citep{1973A&A....24..337S},
$\nu = \alpha C_s^2 / \Omega$, where $C_s$ is the local sound speed, $\Omega$ the
keplerian frequency, and $\alpha$ a free parameter, taken to be $2 \times 10^{-3}$ in this paper.

\subsubsection{Evolution}

The evolution of the gas disc surface density is computed by solving the diffusion equation:
\beq
{d \Sigma \over d t} = {3 \over r} {\partial \over \partial r } \left[ r^{1/2} {\partial \over \partial r}
\tilde{\nu} \Sigma r^{1/2} \right] + \dot{\Sigma}_w(r) + \dot{Q}_{\rm planet}(r)
\label{eqdiff}
.
\eeq
Photoevaporation is included using the model of~\citet{2004MNRAS.347..613V}:
\begin{eqnarray}
\left\{
\begin{array}{lll}
\dot{\Sigma}_w  = 0  &{\rm for}& R < R_g , \\
\dot{\Sigma}_w  \propto R^{-1} &{\rm for}& R > R_g ,
\end{array}
\right.
\end{eqnarray}
where $R_g$ is usually taken to be $5$ AU, and the total mass loss due to 
photo-evaporation is a free parameter. The sink term $\dot{Q}_{\rm planet}$ 
is equal to the gas mass accreted by the forming planets. For every forming planet,
gas is removed from the protoplanetary disc in an annulus centered on the planet, 
and with a width equal to the planet's Hill radius $\rhill = \aplanet \left( { \mplanet \over 3 \mstar } \right)^{1/3} $.

Eq.~\ref{eqdiff} is solved on a grid which extends from the innermost radius of the disc to 1000 AU. At these two points,
the surface density is constantly equal to 0. The innermost radius of the disc is of the order of 0.1 AU, and is taken
from observations (see Table~\ref{table_Andrews} in Sect.~\ref{population}).

\subsection{Protoplanetary disc: solid phase}
\label{solid_disc}

\subsubsection{Planetesimal characteristics}

In our model, we consider two kinds of planetesimals:  rocky  and icy planetesimals. These two kinds
of planetesimals differ by their physical properties, in particular by their mean density (3.2 g/cm$^3$ for the former, 1 g/cm$^3$ for the latter).
Initially, the disc of rocky planetesimals extends from the innermost point in the disc (given by the fourth column of Table \ref{table_Andrews} in Sect. \ref{population}),
to the initial location of the ice line, whereas the disc of icy planetesimals extends from the ice line to the outermost point in the simulation disc.

The location of the ice line is computed from the initial gas disc model, using the central temperature and pressure. The ice sublimation temperature
we use depends on the total pressure. Note that in our model, the location of the ice line does not evolve with time. In particular, no condensation of
moist gas, or sublimation of icy planetesimals is taken into account. Moreover, the location of the ice line being based on the central pressure and temperature,
the ice line is supposed to be independent of the height in the disc. In reality the ice line is likely to be an "ice surface" whose location depends
on the height inside the disc~\citep[see][]{2011Icar..212..416M}. We assume that all planetesimals have the same radius (of the order of 100 m), and that this radius does 
not evolve with time. The extension of our calculations towards a non-uniform and time evolving planetesimal mass function will be the subject of future work.
{Note that the assumed radius of the planetesimals has a strong effect on the resulting formation process of planetary systems. Increasing their radius to a few tens
of kilometers severely decreases their accretion rate and the growth of planets. This effect was discussed in~\citetalias{2013A&A...549A..44F}.}

\subsubsection{Planetesimal surface density and excitation}
 
 The eccentricity and inclination of planetesimals are important since they govern in part the accretion rate of
 solids, and the ability of planets to grow. In the present paper, we follow the approach presented in~\citetalias{2013A&A...549A..44F}
 to compute the r.m.s. eccentricity and inclination of planetesimals as a result of excitation by planets, and
 damping by gas drag. Note that, since more than one planet form in the same disc, planetesimals excited by a planet
 can remain excited when they enter the feeding zone of another planet, modifying their capture probability\footnote{Here,
  we do not include the radial drift of planetesimals. Therefore, planetesimals actually enter the feeding zone of
 another planet, if the feeding zone borders themselves move, as a result of planetary growth and migration. The computation of the
 orbital drift of planetesimals will be the subject of future work.}. 
 
The dynamical state of the planetesimal disc is computed by solving differential equations that describe
the evolution of the excitation and damping rates of their mean eccentricity and inclination.
 As pointed out in~\citetalias{2013A&A...549A..44F}, the memory of the initial value of the planetesimals eccentricity and inclination
 can last for a non-negligible time. We assume here that the initial excitation state of planetesimals is the one resulting from the self-interaction between
 planetesimals alone (cold planetesimals initially). This corresponds to the assumption that planetary embryos (whose mass is $10^{-2} \mearth$) appear
 instantaneously, as a result of e.g. gas-solid interactions~\citep[e.g.][]{2007Natur.448.1022J}.

\subsection{Planetary growth and competition}

\subsubsection{Solid accretion rate}

We start our calculation with a collection of small mass embryos (mass  $10^{-2} \mearth$) which may accrete solids and gas, and
may migrate in the protoplanetary disc. The solid accretion rate of a given embryo is computed following the approach described in~\citetalias{2013A&A...549A..44F},
including the effect of the atmosphere, which enhances the cross-section. The latter is parametrized by the embryo capture
radius, the effective radius of the planets for the accretion of planetesimals, which is computed 
 using the results of~\citet{2003A&A...410..711I}, and is given by
the following implicit equation:
\beq
\rpla = { 3 \rhogas (\rcap) \rcap \over 2 \rhopla }  \left[ {\vrel^2 +  { 2 G m (\rcap) \over \rcap } \over \vrel^2 + { 2 G m (\rcap) \over \rhill } } \right] 
,
\eeq
where {$\rpla$ is the physical radius of the planetesimals,} $\rhogas$ is the density at a distance $\rcap$ from the planet center,   $m(\rcap)$ is the planetary mass inside the sphere
of radius $\rcap$ centered on the planet's center, and $\vrel$ is the relative velocity between the planet and the planetesimals,
which results from their excitation state. Tests have shown that the capture radii obtained with these approximate formula are very
close to the ones obtained by computing the trajectory of planetesimals inside the planetary envelope~\citepalias[see the method described in][]{2005A&A...434..343A}.

\subsubsection{Gas accretion}

The accretion of gas by growing planets is the result of planetary contraction. This is computed by solving the internal structure
equations for the planetary envelope, taking into account as energy source, both the accretion energy of planetesimals,
and the compression work released by the contraction of the planetary envelope. The method is similar to the one presented
in~\citetalias{2013A&A...549A..44F}, to which the reader is referred to for more details. Note that we assume in this model that the {dust} opacity in the planetary envelope
is reduced compared to interstellar values~\citep[see][]{1996Icar..124...62P,2010Icar..209..616M}. For the sake of simplicity and following the
approach of~\citet{1996Icar..124...62P}, we use here a reduction factor of 0.01. We stress however that this value is probably still too high,
and refer the reader to discussions in Mordasini~et~al.~(in~revision) for an in-depth discussion of this effect. The goal of this paper being
the differential effect of multiplicity, the exact value of the opacity reduction factor is of lesser importance.

\subsection{Disc planet interactions}
\label{orbital_evolution}

Disc planet interactions lead to planet migration, which can occur in different
regimes. For low mass planets, not massive enough to open a gap in the protoplanetary disc, migration occurs in type~I~\citep{1997ApJ...482L.211W,2002ApJ...565.1257T,2010MNRAS.401.1950P,2011MNRAS.410..293P}. For higher mass planets, migration is again subdivided into two modes: disc-dominated type~II migration, when the local disc mass is larger than the planetary mass (the migration rate
is then simply given by the viscous evolution of the protoplanetary disc), and planet-dominated type~II migration in the opposite case~\citep[see][]{2009A&A...501.1139M}.
The transition between type~I and type~II migration occurs when~\citep{2006Icar..181..587C}
\beq
{3 \over 4} { \hdisk \over \rhill} + {50 \mstar \over \mplanet \re } = 1
\eeq
where $\hdisk$ is the disc scale-height at the location of the planet, and 
$\re = { \aplanet^2 \Omega \over \nu }$ is the Reynolds number at the location of the planet.
We use in our model an analytic description of type~I migration,
which reproduces the results of~\citet{2011MNRAS.410..293P}, these latter including the effect of co-rotation torque and the fact that
discs can be non-isothermal. A detailed description of this model is presented in Dittkrist~et~al.~(in~prep).

\section{Planet orbital evolution}
\label{nbody}

A key component of our  multiple planetary system model is the calculation of the gravitational interactions between the embryos. This component is not necessary for the description of the evolution of a single planet, but for a multiple planetary system, it can be very important. Gravitational interactions can disturb the orbit of planets and, therefore, increase their collision probability, or inversely they can force them to be trapped in resonances, which can reduce the probability of collision. 
In this section we  describe our method  to calculate the gravitational interactions between the planets and the collision detection, more details are presented in~\citet{2013PhD......Carron}.

\subsection{Equations of Motion}

In the N-body part, we treat each planetary embryo as a point mass. A body is characterized by its position $\vect{x}$, its  velocity $\vect{\dot{x}}$, and its mass $m$. According to Newton's law of universal gravitation, the acceleration of the i-th body  $(\ddot{ \vect{x}})$ can be written as:

\begin{equation}
\label{equ:grav1}
\ddot{ \vect{x_i}}  = - G \sum_{j=0,j\neq i}^{N} {m_j \frac{\vect{x_i}-\vect{x_j}}{\left| \vect{x_i}-\vect{x_j} \right|^3} } ,\hspace{1.0cm} i = 0,1,2,\dots,N.
\end{equation}
where $G$ is the gravitational constant and $N$   the number of planetary bodies. The index $0$ refers to the central star.\\

The equation (\ref{equ:grav1}) can be written for a heliocentric frame. With the heliocentric position $r_i$, defined as:

\begin{equation}
\label{equ:helio}
\vect{r_i} \doteq \vect{x_i} - \vect{x_0} ,\hspace{1.0cm} i = 1,2,3\dots N,
\end{equation}
and with the relation $ \ddot{\vect{r_i}} = \ddot{\vect{x_i}} - \ddot{\vect{x_0}}$, we can write the equation of motion as:

\begin{equation}
\label{equ:grav2}
\ddot{ \vect{r_i}} =-G \left(m_0 + m_i \right) {\vect{r_i} \over r_i^3} - G \sum_{j=1,j\neq i}^{n} {m_j \left\{ \frac{\vect{r_i}-\vect{r_j}}{\left| \vect{r_i}-\vect{r_j} \right|^3} + { \vect{r_j} \over r_j^3 } \right\} }
\end{equation}
with $ i = 1,2,3 \dots N$. This system of  coupled second order differential equations with $3 n$ dimensions is solved using a Bulirsch-Stoer integration scheme.

\subsection{Migration and Damping}

Migration plays a central role during the formation process of planets. Due to the gravitational interaction between the planets and the gas disc, the planets can move  through the disc. The migration pushes the planets inward or outward depending on the properties of the disc and the mass of the planet. We use the method of~\citet{2007A&A...472.1003F} to include the effect of disc planet interaction. The acceleration due to the migration can be written as 
\begin{equation}
\label{equ:migration}
	\vect{a}_m = - \frac{ \vect{v}}{2 t_m},
\end{equation}
  where $t_m$ is the migration timescale defined as $t_m =  - \frac{a} {\dot{a}}$,
    $a$ is the semimajor axis and $v$  the velocity of the body. 
    {This timescale is computed following the work of~\citet{2011MNRAS.410..293P} and depends on the planetary mass, as well as on the local properties of the disc. Details on the migration timescale computation can be found in~\citet{2011IAUS..276...72M} and Dittkrist~et~al.~(in~prep).}
    This equation is valid for small migration forces,  this means  $t_m$ should be much larger than the orbital period.\\

The gravitational interactions of the planets with the gas disc lead to a damping of the eccentricity and of the inclination of the  planets. 
We assume that the eccentricity and inclination damping timescales are similar, and both equal to $1/10$ of the absolute value of the migration timescale.
The ratio between the eccentricity (and inclination) and semi-major axis timescales is very uncertain, and the value
we use there is just a rough order of magnitude estimation. We present in Sect.~\ref{results} our tests to infer the effect of this parameter.

The accelerations caused by the damping  of the eccentricity $\vect{a}_{de}$ and of the inclination $\vect{a}_{di}$ are calculated as follows:

\begin{equation}
\label{equ:damping_e}
	\vect{a}_{de} = -2 \frac{\left( \vect{v}\cdot\vect{r} \right) \vect{r}}{r^2t_e}
\end{equation}
for the eccentricity, and 	
\begin{equation}
	\vect{a}_{di} = - 2 \frac{\left( \vect{v}\cdot\vect{k} \right) \vect{k}}{t_i},
\end{equation}
for the inclination, where $\vect{k}$ is the unit vector $\(0,0,1\)$. Also here, $t_e$ and $t_i$ should be much longer  than the orbital period.

Putting all together we can calculate the total acceleration $\vect{a}_t$ of a planet as:
\begin{equation}
\vect{a}_t = \vect{a}_g + \vect{a}_m + \vect{a}_{de} +  \vect{a}_{di},
\end{equation}  
where $\vect{a}_g $ is acceleration due to the gravitation of the other bodies as described in~Eq.~\ref{equ:grav2} with $\vect{a}_g  \equiv \ddot{ \vect{r_i}}$.

\subsection{Collision Detection}

During the evolution of a planetary system, collisions between planets may occur. We detect collisions by checking, after each time step of the  \textit{N}-body code\footnote{The time step for the
\textit{N}-body is in general much smaller than the time step required to compute planetary growth.}, if two bodies are closer to each 
other than the collision distance $R$, which we define as the sum of both core radii ($R = R_1 + R_2$). The numerical integration has an adaptive time step ($ h $), which ensures that an 
integration with the desired precision  is obtained. In addition, we limit the length of the \textit{N}-body time step to a value smaller than the collision timescale $\tau$, which is calculated as follows: 

\begin{enumerate}
\item  For each pair of planets $ k $,  we approximate the positions $\vect{r_1},\vect{ r_2}$ of the two bodies at the time $t = t_0  + \Delta t$:
 \begin{equation}
 \vect{r_1}(t_0  +  \Delta t) = \vect{x_1}(t_0) + \vect{v_1}(t_0)  \Delta t + \frac{1}{2} \vect{a_1}(t_0)  \Delta t^2
 \end{equation}
 \begin{equation}
 \vect{r_2}(t_0  +  \Delta t) = \vect{x_2}(t_0) + \vect{v_2}(t_0)  \Delta t + \frac{1}{2} \vect{a_2}(t_0)  \Delta t^2,
 \end{equation}
 where $\vect{x_1}(t_0),\vect{x_2}(t_0)$ are the positions,  $\vect{v_1}(t_0),\vect{v_2}(t_0)$ the velocities, and $\vect{a_1}(t_0),\vect{a_2}(t_0)$ the accelerations at the time  $t_0$.
 \item We define $\tau_k $ as the minimum real solution for the collision timescale of the equation:
 \begin{equation}
  \left( \vect{r_1}(t_0 +  \Delta t) -  \vect{r_2}(t_0 +  \Delta t)\right)^2 = R^2.
 \end{equation}
 With the substitution 
\begin{equation}
\Delta \vect{x} = \vect{x_1}(t_0) - \vect{x_2}(t_0) 
\end{equation}
\begin{equation}
\Delta \vect{v} = \vect{v_1}(t_0) - \vect{v_2}(t_0) 
\end{equation}
\begin{equation}
\Delta \vect{a} = \vect{a_1}(t_0) - \vect{a_2}(t_0)
\end{equation}
we get 

\begin{eqnarray}
\begin{array}{r}
 \frac{1}{4} \left(\Delta \vect{x} \right)^2 \Delta t^4 + \Delta \vect{v} \Delta \vect{a} \Delta t^3 + \left( \left(\Delta \vect{v} \right)^2 + \Delta \vect{x} \Delta \vect{a} \right) \Delta t^2   +  \\
2 \Delta \vect{x} \Delta \vect{v} \Delta t  + \left(\Delta \vect{x}\right)^2 - R^2 = 0 
\end{array}
\end{eqnarray}

\item
We are only interested in solutions with $ \tau_{k} < h $, $ h $ being the time step of the \textit{N}-body integrator. If the distance between the two bodies $ d $  at the time $ t_0 $:
\begin{equation}
d = |\vect{x_1}(t_0) -  \vect{x_2}(t_0) | - R,
\end{equation}
is larger than the maximum change of the distance $ \Delta d_{max} $,
\begin{eqnarray}
\begin{array}{lcl}
\Delta d_{max}&  = & \max \left(| \vect{r_1}(t_0 + \tilde \Delta t) -  \vect{r_2}(t_0 + \tilde \Delta t)| \; \forall \tilde \Delta t \in [t_0,t_0 + \Delta t] \right) \nonumber \\
& = &  \max \left(| \Delta \vect{v}(t_0 + \tilde \Delta t) + \frac{1}{2} \Delta \vect{a}(t_0 + \tilde \Delta t)^2| \; \forall \tilde \Delta t \in [t_0,t_0 + \Delta t] \right) \nonumber
\end{array}
\end{eqnarray}
then, obviously, no real solution exists with $ \tau_{k} < \Delta t $.\\
Calculating  $ \Delta d_{max} $ is not trivial, however we can easily find a maximum limit of  $ \Delta d_{max} $ using the triangle inequality:
\begin{eqnarray}
\begin{array}{lcl}
  \Delta d_{max}  & = & \max \left(| \Delta \vect{v}(t_0 + \tilde \Delta t) + \frac{1}{2} \Delta \vect{a}(t_0 + \tilde \Delta t)^2| \; \forall \tilde \Delta t \in [t_0,t_0 + \Delta t] \right) \nonumber \\
& \le &  \max \left( \Delta \vect{v}| (t_0 + \tilde \Delta t) + | \frac{1}{2} \Delta  \vect{a}|(t_0 + \tilde \Delta t)^2 \; \forall \tilde \Delta t \in [t_0,t_0 + \Delta t] \right) \nonumber \\
& = & | \Delta \vect{v}| (t_0 + \Delta t) + \frac{1}{2} |\Delta \vect{a}|(t_0 + \Delta t)^2 
\end{array}
\end{eqnarray} 
This leads to  $ \Delta d_{max} \le  | \Delta \vect{v}| (t_0 + \Delta t) + |\frac{1}{2} \Delta \vect{a}|(t_0 + \Delta t)^2 $

\item 
We define $ \tau $ as the minimal $ \tau_{k}, \mathrm{where\;} k = 1...N(N-1)/2 $
\end{enumerate}
Our model is very similar to the one described in~\citet{2000Icar..143...45R}. The only difference is that we use a second order Taylor series for the approximation of the position instead of a first order one. Therefore the method is more accurate but has the disadvantage to bring out a fourth order equation instead of one of second order.  
\\

\subsection{Collision handling} 

After each \textit{N}-body time step we check for collisions. If one (or more) is found, we merge the colliding bodies: the collisions are treated as fully inelastic. 
This means that we remove the less massive body and   add its mass to the more massive one. We also change the position  and velocity of the more massive one so that the total momentum of the centre of mass is conserved.

When two planets merge, the resulting planet has a core mass that is the sum of the two core masses. 
The envelope mass of the new planet is calculated as follows: we compute the collision energy and compare it to the binding energy of the more massive planet's envelope. 
If the former is the largest, the envelope of the planet is ejected, otherwise, it is conserved.  In the case where both planets are massive, 
with large envelopes each, our treatment is not accurate. However, we do not expect such collisions to occur, as these planets would probably be captured in mutual 
resonances and would not collide.  For details on the accretion of a solid embryo by a planet with an envelope we refer the reader to~\citet{2012A&A...538A..90B}.  

\section{Competition for gas and solids accretion}
\label{competition}

The planet's feeding zone is assumed to extend to $4 \rhills$ on both sides of the planet. In case a planet has an eccentricity,
the feeding zone extends from $a_{\rm min} -  4 \rhills$ to $a_{\rm max} + 4 \rhills$, where $a_{\rm min}$ and $a_{\rm max}$ are the periastron
and apoastron.  An important effect in our model is the treatment of the feeding zones of planets when they overlap. Indeed, we assume in our model,
as was done in~\citet{1996Icar..124...62P}~and~\citetalias{2005A&A...434..343A} for example, that the planetesimal surface density in a planet's feeding zone is uniform. As a consequence,
if the two feeding zones of two different planets overlap, the planetesimal surface density in the global feeding zone itself is constant. This
has a number of potentially important consequences:
\begin{itemize}
\item since a planet's envelope depends upon its luminosity, {which itself depends} on the planetesimal accretion rate, and therefore on the
planetesimal surface density, when two feeding zones overlap, the internal structure of the two planets is no more 
independant. 
\item two planets sharing their feeding zones compete for the accretion of planetesimals. Interestingly enough, this does not  necessarily result
in a reduced solid accretion rate. In general, one planet is favored (its accretion rate is increased compared to the corresponding isolated
situation), whereas the other one will grow slower. This results simply from the fact that if the two planets compete for the accretion of the same
planetesimals (which should result in a decrease of the solid accretion rate for both planets), they also have access to a much larger region
of the disc (the union of the two feeding zones of the two planets).
\end{itemize}

Numerically we proceed as follows. When two feeding zones overlap, we consider that they merge into a big one,  its inner limit ($a_\mathrm{inner}$) 
being the minimum of the two inner boundaries of the separated feeding zones, and the outer limit ($a_\mathrm{out}$) the maximum of the two outer ones. 
The surface density of the new feeding zone is considered to be uniform, i.e. the solids surface density of the region is integrated to obtain the total mass
 which is then divided by the surface of the feeding zone. 

\begin{figure}
  \center

  \includegraphics[height=0.5\textheight]{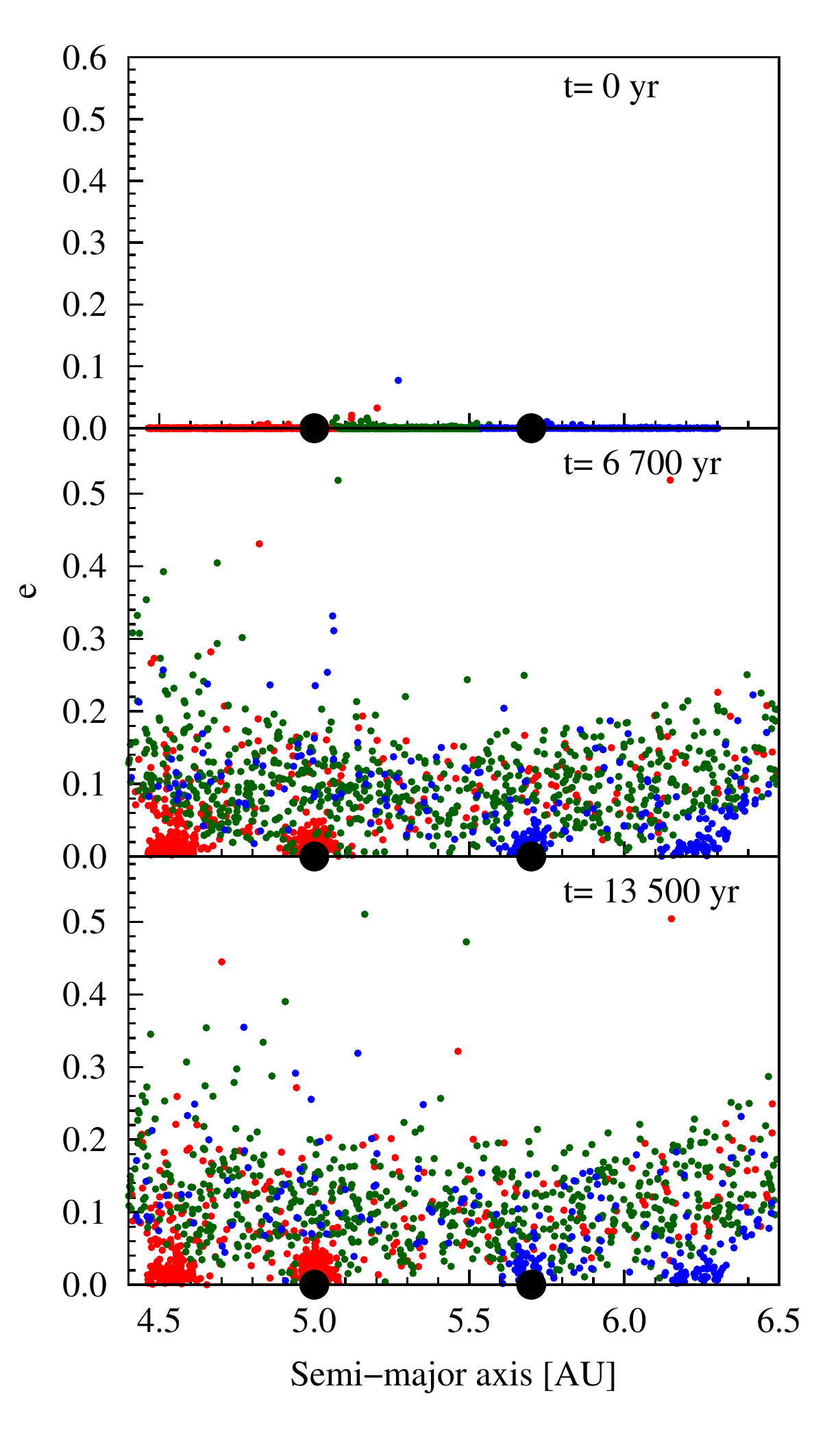}
    
  \caption{Evolution of the eccentricity (y axis) and semi-major axis (x axis, in AU) of a set of test particles under the influence of two planets without gas drag. Particles are coloured according to their initial location:
  red for particles belonging to the feeding zone of the innermost planet, blue for particles belonging to the feeding zone of the outermost planet, and green for
  particles belonging to both feeding zones. The two 10 M$_\oplus$ planets are located at 5 AU and 5.6 AU. The first panel shows the initial conditions, the second panel is the state
  after 600 orbital times of the innermost planet, and the last panel is the state after 1200 orbital times of the innermost planet. As can be seen, test particles are very efficiently mixed in
  the global feeding zone on a short timescale. Only planetesimals located in co-rotation resonance with one of the two planets are scattered on a longer timescale.}
  \label{fig_Samuel}
\end{figure}

To check if our prescription is a good approximation of reality, we performed $N$-body calculations, considering two planets and a set of test particles. 
Fig. \ref{fig_Samuel} shows the semi-major axis and eccentricity of the test particles at different times.  Particles have different colours 
depending on their initial location (see caption of Fig. \ref{fig_Samuel}). As can be seen in the figure, the planetesimals are indeed very efficiently mixed,
 resulting in a quasi-uniform surface density (and eccentricity) in a global feeding zone, thus validating our approach.

Planets also compete for the accretion of gas. Indeed, there is a maximum mass a planet can accrete during a given time step. This mass is the sum of the mass already present
in the planetary gaseous feeding zone (which is assumed to extend to one Hill's radius on both sides of the planetary location),
and the mass that can enter in the feeding zone during the time step, as a result of viscous transport~\citepalias[see][]{2013A&A...549A..44F}. Since,
as already mentioned above, the gas accreted by forming planets is removed from the protoplanetary disc, the mass reservoir
as well as the viscous transport are modified when a planet accretes a large amount of gas (in particular during the runaway gas accretion).
These modifications can in turn modify the maximum mass another planet can accrete. 

\section{Results}
\label{results}

\subsection{Formation of a 10-planet system}
\label{formation_10planets}

Before considering a population of planets, we study an example of a 10-planet system formation model. We have considered
three models with the same initial conditions. The first one assumes that only one planet forms in the protoplanetary disc (this means that
we ran 10 independent models, varying only the initial location of the planetary embryo, which are taken as the same initial location of planets
in the 10-planet case). The second one takes into account the competition
between planets for solids and gas accretion, the excitation of planetesimals by all the planets, but not the gravitational interactions between planets. 
The third model takes into account both the competition for accretion and the gravitational interactions between forming planets: the orbits of 
planets are computed using the \textit{N}-body described above.
{For the three models, the initial surface densities of gas and solids were taken to be 140 g/cm$^2$ and 6 g/cm$^2$ at 5 AU, 
corresponding to 730 g/cm$^2$ and  8 g/cm$^2$ respectively at 1 AU.}

Note that in the second model, we use a very simple prescription for planetary collisions: as soon as two planets have the same semi-major axis, 
the smallest one is either accreted or ejected by the biggest one.  The ratio of ejection to accretion probability is assumed to be the same as for planetesimals~\citepalias[see][]{2005A&A...434..343A}. This is of course a very simplified model, but its only purpose is to emphasize the differences with the third model.


\begin{figure}
  \center
  \resizebox{\hsize}{!}{
    \includegraphics[width=0.25\textheight,angle=0]{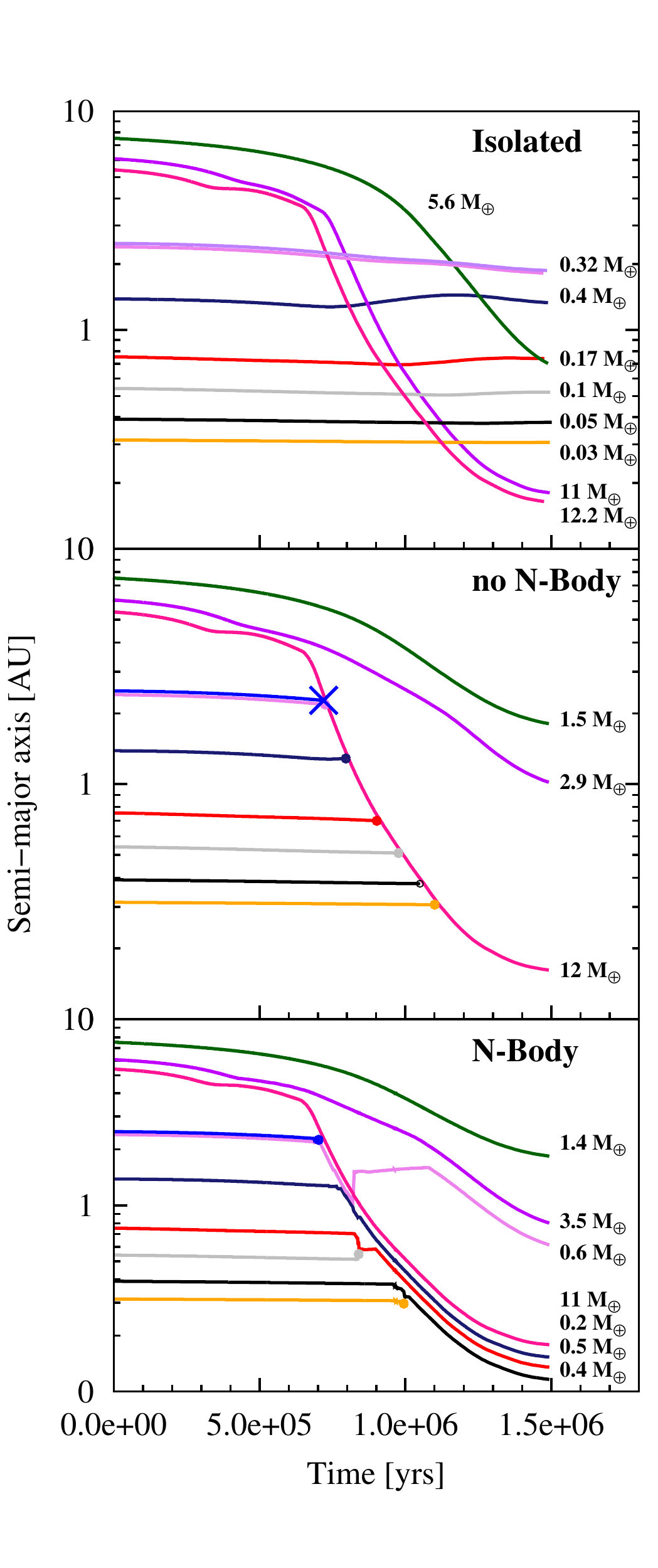}}
  \caption{Three models for the formation of 10 planets. {\bf Top}: formation of 10 independent planets (each of them growing in an identical disc). {\bf Middle}: the competition for
  gas and solid accretion, as well as the excitation of planetesimals by all planets, is taken into account. The gravitational interactions between planets are not
  included. The cross indicates an ejected planet, the big dots indicate collision between planets {\bf Bottom}: Full model, including the competition for gas and solid accretion, 
  the excitation of planetesimals by all planets, and the gravitational interactions
  between planets.}
  \label{10planets}
  \end{figure}

The first model led to a system with 10 planets inside 2\,AU, and with masses ranging from 0.03$\,\mearth$ to $\sim 12\,\mearth$ (Fig.~\ref{10planets}, upper panel).

The results using the second model are presented in Fig. \ref{10planets}, middle panel. The planet that grows more massive is located initially at 5.4\,AU, a privileged place in
terms of abundance of solids and size of the feeding zone\footnote{The ice line is located at 2.8\,AU, and the planet starting at 5.4 AU is the planet
starting both outside the ice-line, and closer to the star than the others.}. During its inward migration it encounters seven other smaller planets, one of 
which is ejected while the others are accreted. The total mass that the planet gains by accreting
other embryos is 0.63$\,\mearth$. In the system, the two outer planets that never suffer from any encounter remain, so in total
the system ends up with 3 planets out of the initial 10. 
The final masses are 12, 2.9 and 1.5$\,\mearth$, the final locations
being 0.16 AU, 1 AU and 1.8 AU respectively. In the case of the most massive planet, its formation is almost identical as if
it were growing as an isolated planet. The inner embryos, most of which are accreted, do not favour or slow down the growth
of the planet. 

As an illustration of the importance of planetary interactions, we note that the final mass of the outermost planet 
 turns out to be 5.6 $\mearth$ and its final location 0.75 AU (if only one planet is considered), to be compared with
 1.5 $\mearth$ at $\sim 2$ AU in the second model. Indeed, in the multi-planet
case accretion is largely reduced when the planet enters regions of the disc already visited by other embryos (and therefore with
less material available) and by the fact that random velocities of the (fewer) available planetesimals in these regions are higher
due to the perturbations of the other protoplanets. Note however that the reduction of solid accretion could also reduce the critical
mass, and therefore enhance gas accretion. For this process to happen, however, the reduction of solid accretion must occur for a planet that
is already quite massive, a situation that is not encountered in this simulation.

\begin{figure}
  \centering
  \includegraphics[width=0.35\textheight,angle=0]{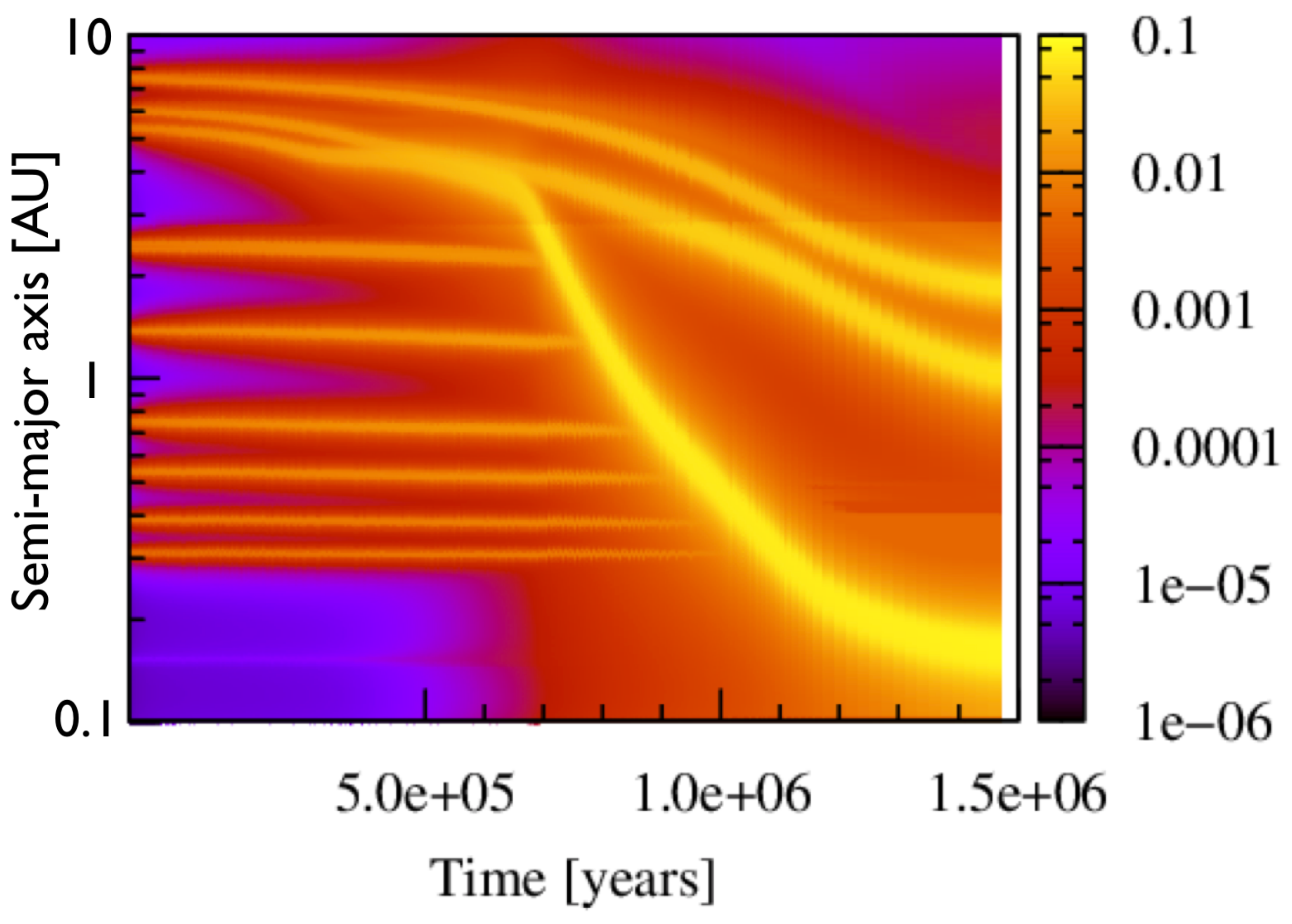}
  \caption{Eccentricity of planetesimals in the disc as a function of time (x-axis) and semi-major axis (y-axis).  The colour bar indicates the eccentricity values. In this model, the
  competition for gas and solid accretion, as well as the excitation of planetesimals by all growing planets, are included. This corresponds to the middle panel of Fig. \ref{10planets}.}
  \label{10planets_e}
\end{figure}

To illustrate the excitation of planetesimals by the 10 planets, Fig.  \ref{10planets_e} shows the eccentricity of planetesimals in the disc as a function of semi major axis and
time in the second model. Clearly, at the position of the protoplanets the eccentricity is the largest. As the planets grow they perturb the disc to a
great extent. {In addition, as the disc dissipates, the damping effect of the gas drag decreases, which results in an overall larger excitation of planetesimals.}

Using the third model, we obtain a different system (see  Fig. \ref{10planets}, bottom panel). 
As we can see, in this case only three planets are accreted by the most massive planet during its inward migration,
so the final number of planets in this system is 7. Planets that before were considered to be accreted or ejected, survive in the
system due to resonance trapping. Also, orbit crossing is possible without the loss or the ejection of the planet. When the gravitational
interactions are not considered, small planets that are in the inner part of the disc are usually swept out by a more massive,
inward migrating planet. However, as we can see from this example, this approximation underestimates the amount of these
small, close-in planets. Therefore, accurate formation of planetary systems should account for the gravitational interactions
of the protoplanets: To get the proper orbital configurations in planetary systems, N-body calculations are mandatory.


\subsection{Planet population}
\label{population}

We now present the effect of multi-planetary formation at the population level, by comparing two identical models, one assuming only
one planet growing in each protoplanetary disc, the second one assuming that 10 planetary embryos grow in each of the discs.
We stress that it is not our goal in the present paper to reproduce the observed population
of planets, but rather to study the differential effect of having more than one planet forming in a system, using parameters that lead to populations
not totally at odd with observations.

\subsubsection{Initial conditions}

The initial conditions are given by the characteristics of a protoplanetary disc, and the ensemble of planetary embryos,
whose initial mass and semi-major axis are computed in the following way. The starting location
of the planetary embryos is selected at random, using a probability distribution uniform in log. It ranges
from 0.1 AU to 20 AU. We moreover impose that two planetary embryos should not start closer than 10 times their mutual Hill radius.
The initial mass of the planetary embryos is assumed to be equal to $10^{-2} \mearth$. 

 The initial gas disc surface density profiles we consider are given by:
\beq
\Sigma = (2 - \gamma) { \mdisk \over 2 \pi \ac ^{2-\gamma} \rzero^\gamma } \left( {r \over \rzero} \right)^{-\gamma} \exp \left[ - \left( {r \over \ac} \right) ^{2-\gamma} \right]
,
\eeq
where $\rzero$ is equal to 5.2 AU, and $\mdisk$, $\ac$, $\gamma$ are derived from the observations of \citet{2010ApJ...723.1241A}. For numerical reasons, the innermost disc radius, $\rin$ is taken at 0.05 AU, and differs
in some cases from the one cited in \citet{2010ApJ...723.1241A}. Although \citet{2010ApJ...723.1241A} derive  a value for the viscosity
parameter $\alpha$, we assume for simplicity that the viscosity parameter is the same for all
the protoplanetary discs considered. Using a different $\alpha$ parameter will be the subject of future work. We assume that the mass
of the central star is $1 \msun$.

\begin{table}
\caption{Characteristics of disc models}
\begin{tabular}{lcccc}
\hline\noalign{\smallskip}
disc & $\mdisk$ ($\msun$) & $\ac$ (AU)  & $\rin$ (AU) & $\gamma$ \\
\noalign{\smallskip}
\hline\noalign{\smallskip}
1 & 0.029 & 46 & 0.05  & 0.9 \\
2 & 0.117 & 127 & 0.05  & 0.9 \\
3 & 0.143 &198 & 0.05 & 0.7 \\
4 & 0.028 & 126 & 0.05  & 0.4 \\
5 & 0.136 & 80 & 0.05  & 0.9 \\
6 & 0.077 & 153 & 0.05  & 1.0 \\
7 & 0.029 & 33 & 0.05 & 0.8 \\
8 & 0.004 & 20 & 0.05  & 0.8 \\
9 & 0.012 & 26 & 0.05  & 1.0 \\
10 & 0.007 & 26 & 0.05  & 1.1 \\
11 & 0.007 & 38 & 0.05  & 1.1 \\
12 & 0.011 & 14 & 0.05  & 0.8 \\
\hline
\end{tabular}
\label{table_Andrews}
\end{table}

As in \citet[][\citetalias{2009A&A...501.1139M,2009A&A...501.1161M} in the following]{2009A&A...501.1139M,2009A&A...501.1161M}, the planetesimal-to-gas ratio is assumed to scale with the metallicity of the central star, {with a ratio
of 0.04 for solar metallicity. For the disc models we consider, this corresponds to solid surface densities ranging from 0 to 10 g/cm$^2$ at 5.2 AU, with a long tail distribution
extending up to 50 g/cm$^2$}.
For every protoplanetary disc we consider, we therefore select at random the metallicity of a star from a list of 
$\sim 1000$ CORALIE targets~(Santos~priv.~comm.).
Finally, following \citet{2009AIPC.1158....3M}, we assume that the cumulative distribution of disc lifetimes decays exponetially
with a characteristic time of 2.5 Myr. When a lifetime $\tdisk$ is selected, we adjust the photoevaporation rate so that the
protoplanetary disc mass reaches $10^{-5} \msun$ at the time $t = \tdisk$, when we stop the calculation.

\subsubsection{Mass \textit{versus} semi-major axis diagrams}

The number of planetary embryos we consider in each protoplanetary disc is a free parameter. In order to ease the comparison between the two computations, the total number of planets in each case
is similar (at least at the beginning of the calculation): we have considered 500 systems with 10 planets, and $\sim 5000$ systems with only one planet. The initial locations
of planets, in the two cases, are statistically the same, but, as opposed to what was presented in Sect. \ref{formation_10planets}, the starting location of planets in the 1-planet
case are not exactly the same as in the 10-planet case.  

\begin{figure}
  \center

  \includegraphics[width=0.25\textheight,angle=-90]{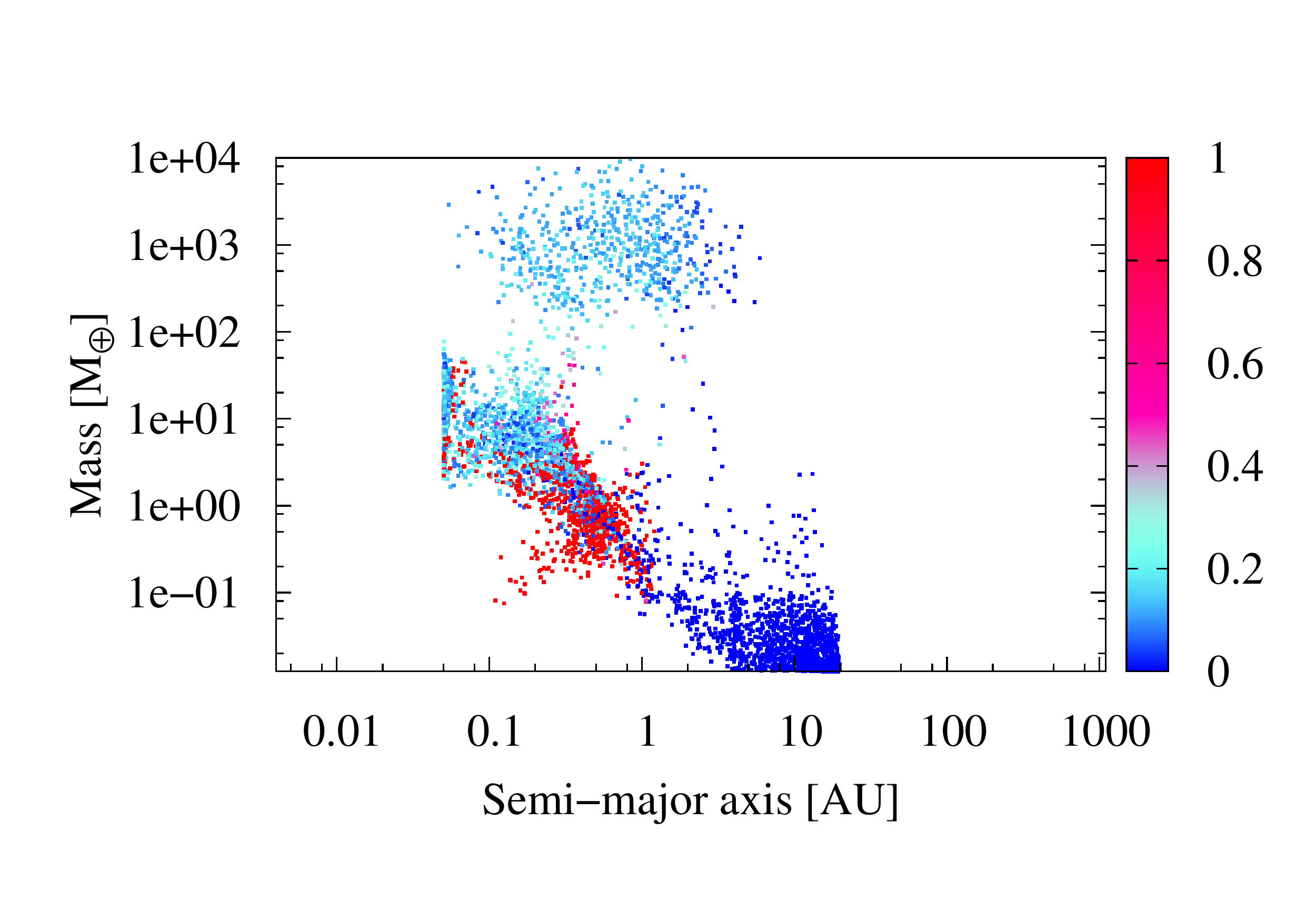}
    
  \caption{Mass \textit{versus} semi-major axis diagram for a population of planets, assuming one planet grows in each disc. The color code shows the 
 fraction of rocky planetesimals accreted by the planet. Planets whose core is the result of the accretion of rocky
 planetesimals are in red, whereas planets whose core has been made by the accretion of icy planetesimals are in blue. 
  The total number of point is 4936. Planets in the vertical line at 0.05 AU are planets that
 reached the inner boundary of the computed disc. Their fate, if the computational domain were extended to lower semi-major axis, is  uncertain.
  They could continue migrating toward the central star, and be accreted, or could stop their migration somewhere in the inner disc cavity. }
  \label{am_single}
\end{figure}

\begin{figure}
  \center

  \includegraphics[width=0.25\textheight,angle=-90]{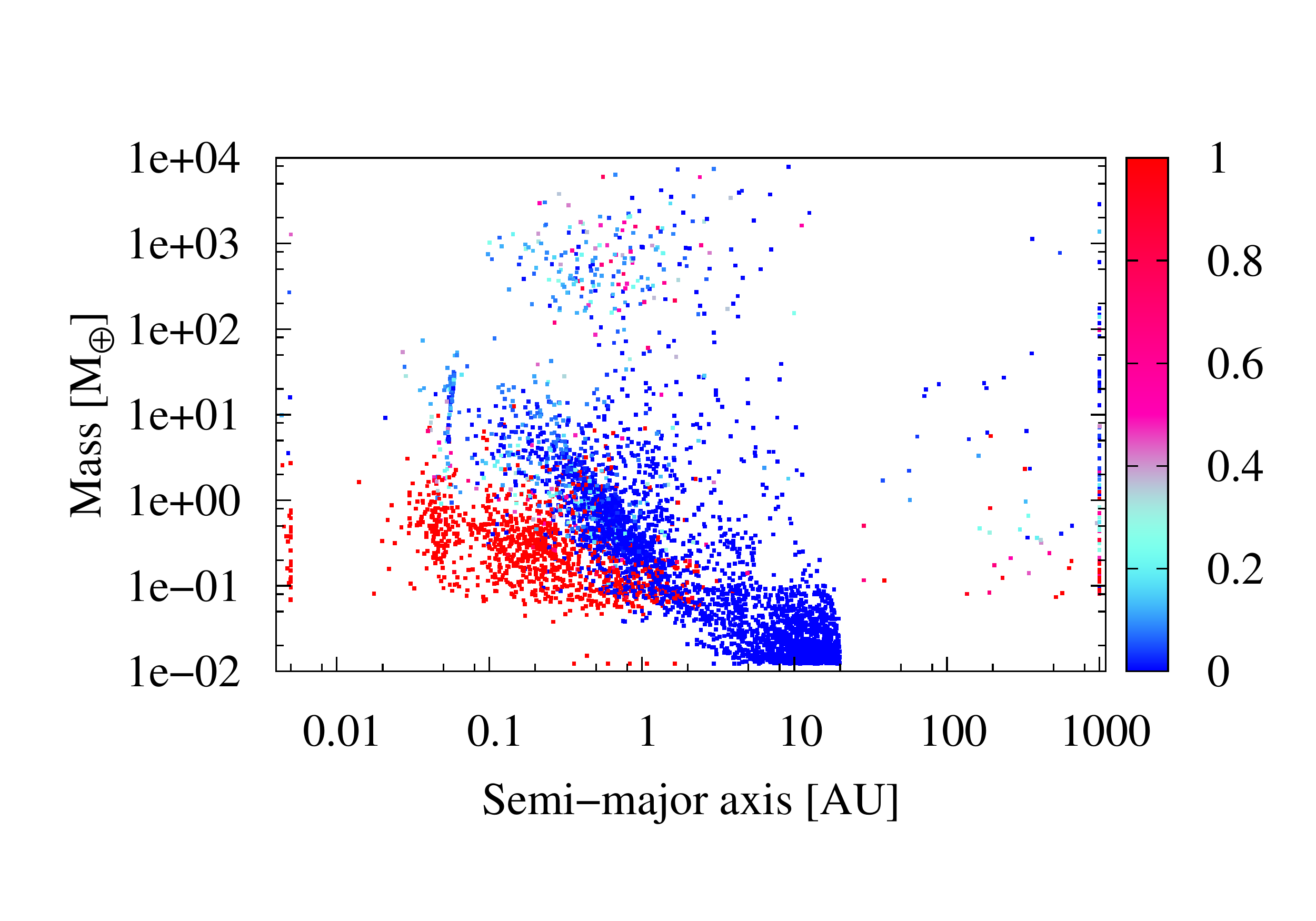}
    
  \caption{Same as Fig. \ref{am_single}, but assuming now that 10 planetary embryos are growing and migrating in every
  protoplanetary disc. The number of points is 5010. Planets on the vertical line at 1000 AU are planets ejected from the system. Their mass represents
  their mass at the time of ejection. Planets on the vertical line at 0.005 AU are planet that have collided with the central star. Note that we do not include in
  our models planet-star interactions that could modify the orbital evolution of planets in the innermost parts of the disc.}
  \label{am_systeme}
\end{figure}

Fig. \ref{am_single} shows the mass \textit{versus} semi-major axis diagram of synthetic planets, in the case where only one planet forms in the system (case 1).  
The color code is related to the composition of the planetary core, which itself is the result of the accretion of different kinds of planetesimals (icy planetesimals or rocky planetesimals).
Blue points are for planets whose core results entirely from the accretion of icy planetesimals, whereas red points are for planets whose core results
from the accretion of only rocky planetesimals. Fig. \ref{am_systeme} presents the same results, but in the case of 10 planets per system.


Comparing the two diagrams (Fig. \ref{am_single} and \ref{am_systeme}) , it appears clearly that not all planets (in terms of mass and semi-major axis) are affected in the same way by the
presence of other bodies.  In particular,  the sub-population of massive planets does not seem to be affected as much,
although planets in the 10-planet case are slightly less massive. Another interesting difference is that planets in the one-planet case are located closer to the
central star, still in the same mass domain. The origin of both differences is the competition between planets forming in the same disc. As planets compete
for the accretion of solids, their growth is delayed. They start to migrate later in the disc lifetime, and start to accrete gas in a runaway mode at a later time.
As a consequence, their final location is somewhat further out, compared to the one-planet case, and their mass is smaller (note that the mass of the planets
is plotted using a logarithmic scale, which decreases the visual difference between the two populations).

In the sub-population of low-mass planets, in particular close to the central star, the effect
of multiplicity is very important. In the 10-planet case, a population of close-in Earth- to Super-Earth mass planets appear, whereas this region
is empty in the case of 1-planet systems. This difference originates from the gravitational interactions between planets in the same system: At a fraction
of an AU from the central star, the mass of solids (these planets are made almost totally from solids) is not large enough to grow a planet of a few Earth masses, at least
for the disc masses we consider here. On the other
hand, disc planet angular momentum exchange alone (leading to migration) is not large enough for these planets to move planets from the outside toward
this region. As a consequence, planets at these distance are either less massive than the Earth, or more massive than $\sim 10 \mearth$. In the case of a multi-planetary
system, planets interact gravitationally with another member of the same system, which itself is massive enough to migrate appreciably. As a consequence, an inner,
low mass planet, can be pushed by resonant interaction toward the inner parts of the protoplanetary disc. Note however that this does not imply that the different planets are
in mean motion resonance at the end of the protoplanetary disc lifetime. Indeed, migration depending on the planetary mass, a mean motion resonance can be broken
during a later phase of disc evolution.

A third sub-population that is notably different between the two cases, is the population of planets below 0.05 AU, at all masses. The difference again stems
from the resonant interaction between planets. In the one-planet case, since the protoplanetary disc is assumed to extend down to 0.05 AU only, migration
ceases for planets below this radius. In the 10-planet case, on the other hand, planets can suffer resonant interaction and enter the innermost parts of the disc.
It should be noted, however, that this difference depends strongly on the adopted value of the disc inner cavity radius. 

A fourth difference is related to planets located at large distances from their central star. Obviously, since the initial location of the planets is assumed to be smaller than 20\,AU, planets in
the one-planet case are all located in the inner regions of the disc (although planets can {migrate outward during some phases of their formation}, they globally
terminate their migration at a position closer to the star than the initial one). In the 10-planet case, gravitational interactions between planets can lead to the scattering
of planets either towards the outer regions of the disc (few hundreds of AU), or ejecting them from the system alltogether  (the outer boundary of the system is assumed to be at 1000 AU). 
Some of the planets ejected from the inner regions of the system, but still bound to the star, are quite massive and could be compared with planets detected by direct imaging~\citep[e.g.][]{2008Sci...322.1348M,2008Sci...322.1345K,2009A&A...493L..21L}.

Interestingly enough, there seems to be a lack of massive
planets at "intermediate" distance (50-100 AU). It is not presently clear if this is due to low statistic effects, or if this is a real effect (for example due to the initial
location of planetary embryos, assumed to be below 20 AU). In addition, microlensing surveys have recently claimed the discovery of  a large population of
massive free floating planets. As can be seen in Fig.~\ref{am_systeme}, planets with very large masses can indeed be ejected from the system during the formation.
We finally note the results presented in Fig.~\ref{am_systeme} correspond to the state of the system at the time the gas disc vanishes. We do not include in these calculations
the long term evolution of planetary systems. Such a study, and the study of the resulting eccentricity evolution of planets, is beyond the scope of this paper,
and will be considered in a forthcoming work (Pfyffer~et~al.,~in~prep). Such effects could increase the number of ejected planets, and as a consequence the number
of expected free floating planets.

Finally, a last difference is in the composition of planets, in particular in the Super-Earth mass domain. Indeed, planets in this mass range
are notably richer in volatile elements in the 10-planet case, compared to the 1-planet case. The origin of this difference is again related to a modified migration
of planets, thanks to gravitational interactions. The ice line is located in our disc at few AUs from the central star. As a consequence, planets below 1 AU are
in general devoid of volatiles, except if they are massive enough to have migrated significantly. On the other hand, in the multi-planet
case, low mass planets, starting their formation in the cold parts of the disc (where planetesimals are volatile rich) can be pushed to the volatile poor regions of
the disc by another external and more massive planet. While this effect is likely to be quantitatively modified including the orbital drifting of planetesimals as a result
from gas drag, this effect should qualitatively remain present in more detailed models.

\begin{figure}
  \center
  \includegraphics[width=0.35\textheight]{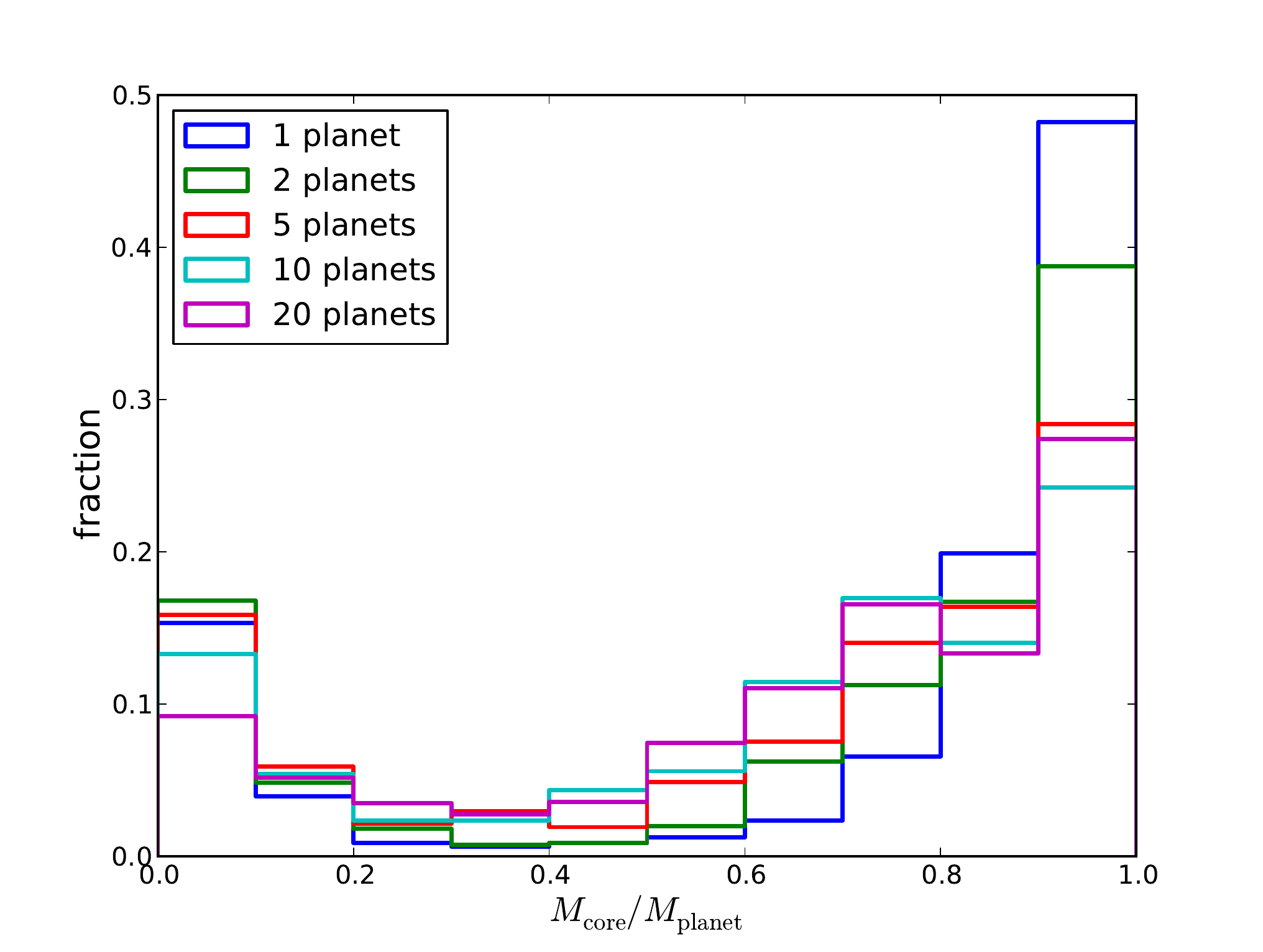}
  \caption{Distribution of the fraction of heavy elements $\mcore /  \mplanet$ for the same simulations, differing only by the assumed number of planetary embryos per system. Only planets
  more massive than $1 \mearth$ have been considered in these distributions.}
  \label{histo_compo}
\end{figure}

{Also related to the composition of planets, we have compared the mean metallicity of planets in the different cases. For this, we have plotted in Fig. \ref{histo_compo}  the histogram
of $\mcore / \mplanet$ for the different cases we have considered (including the calculations with 2, 5 and 20 planetary embryos, see below). Interestingly enough, the
mean heavy element fraction increases for planets forming in systems. This effect can be explained as follows: planets forming in systems acquire their mass on a longer
timescale (compared to single planets). As a consequence, they reach the critical mass later, and have less time (until the gas disc dissipates) to accrete gas.}

{Note that in both populations, models predict the existence of a population of low mass objects at distances between a few AU and 20\,AU.
These represent planetary embryos that have not managed to grow larger than a fraction of an Earth mass. Their outermost location (20\,AU) is simply the effect of
the assumed initial location of planetary embryos (which only extends to 20\,AU). Their innermost location corresponds to places in the disc where the solid accretion
rate of solids is so small that planetary embryos do not grow noticeably during ths lifetime of the gas disc.}

\section{Discussion and conclusions}
\label{discussion}

\subsection{Effect of multiplicity}

In the previous section, we have presented the differential effects of considering the formation of more than one planet in the same
disc. However, there are parameters that could potentially have important effects on the results. In particular, we considered
the growth and migration of 10 planetary embryos, and one may wonder what would result if this number was changed. In addition, we
have mentioned earlier that the timescale for damping of the planet's eccentricity and inclination is theoretically poorly
known.

\begin{figure}
  \center
  \includegraphics[width=0.25\textheight,angle=-90]{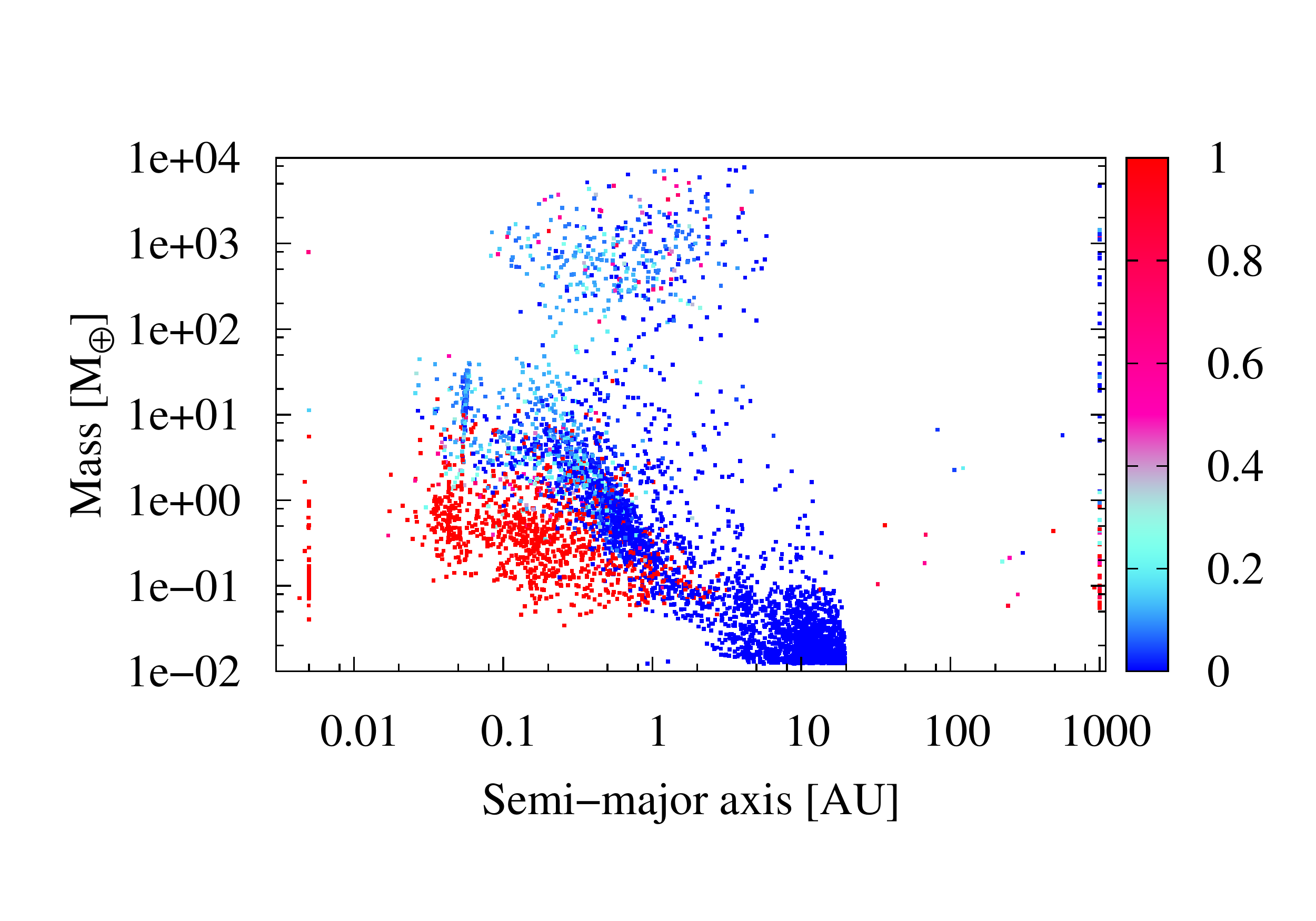}
  \includegraphics[width=0.25\textheight,angle=-90]{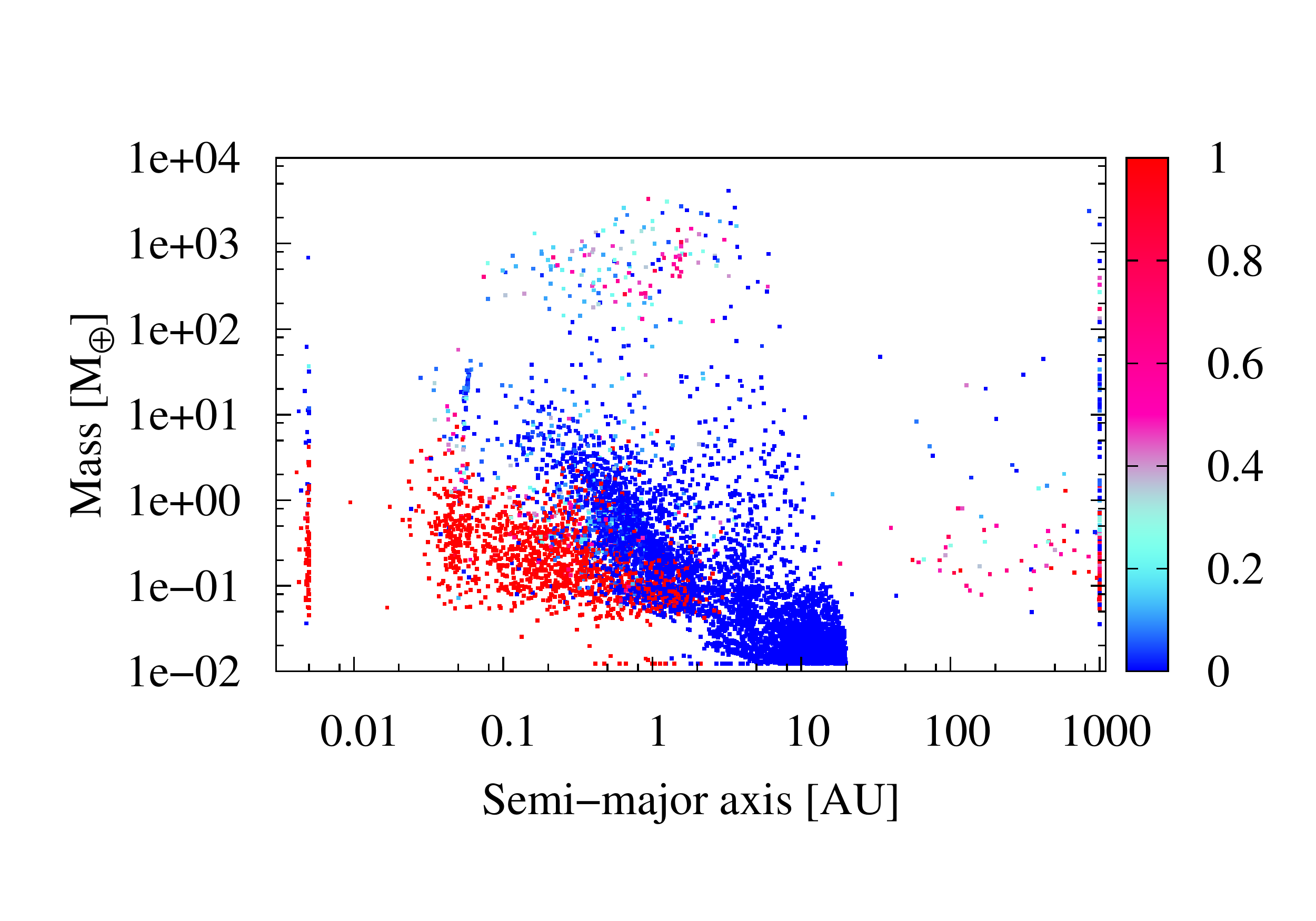}
  \caption{Same model as in Fig. \ref{am_systeme}, except that 5 (top) or 20 (bottom) embryos are assumed to form in the same protoplanetary disc. The number of points are 4875
  and 5000 respectively.}
  \label{am_CD2159}
\end{figure}

To check on the sensitivity of our results to the number of starting embryos used, we initiated a set of additional simulations
with 2, 5, and 20 embryos. The resulting masses and semi-major axis diagrams
for the simulations with 5 and 20 planets are depicted  in Fig. \ref{am_CD2159}. 
As can be seen on the two figures, and comparing with Fig. \ref{am_systeme}, the effect of the number of embryos, at least on the mass \textit{versus} semi-major axis diagram, is modest: the
global structure is similar. One can note, however, that the population of planets at large distances (beyond 50 AU) is larger in the case of 20 planetary embryos. Moreover, the population
of massive planets (larger than Jupiter) is smaller in the 20 planets case. Finally, a population of intermediate planets (from Super-Earth to Neptune mass) at {a few AU appears in} the 20 planets case.


\begin{figure}
  \center

  \includegraphics[width=0.3\textheight]{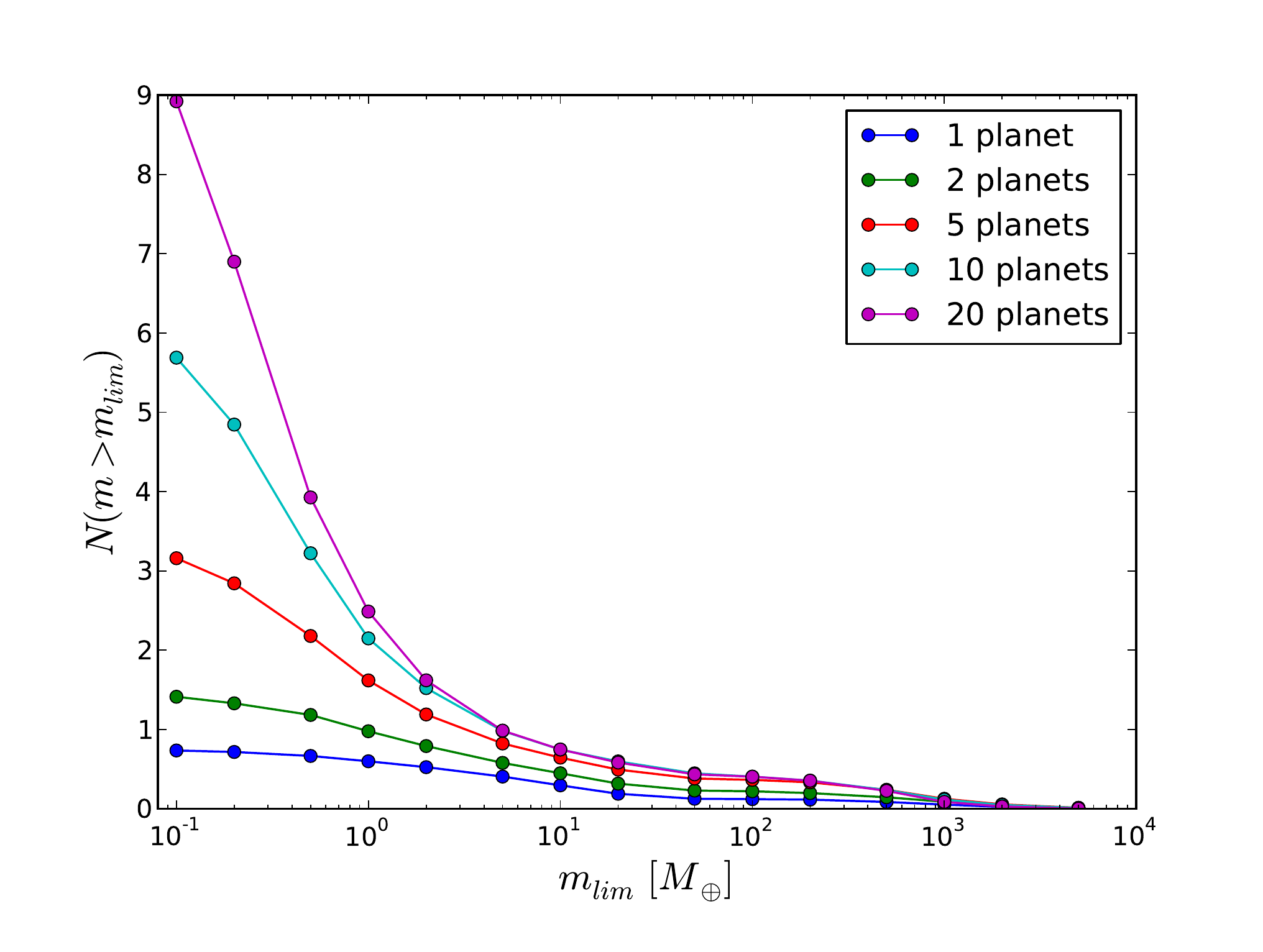}
  \includegraphics[width=0.3\textheight]{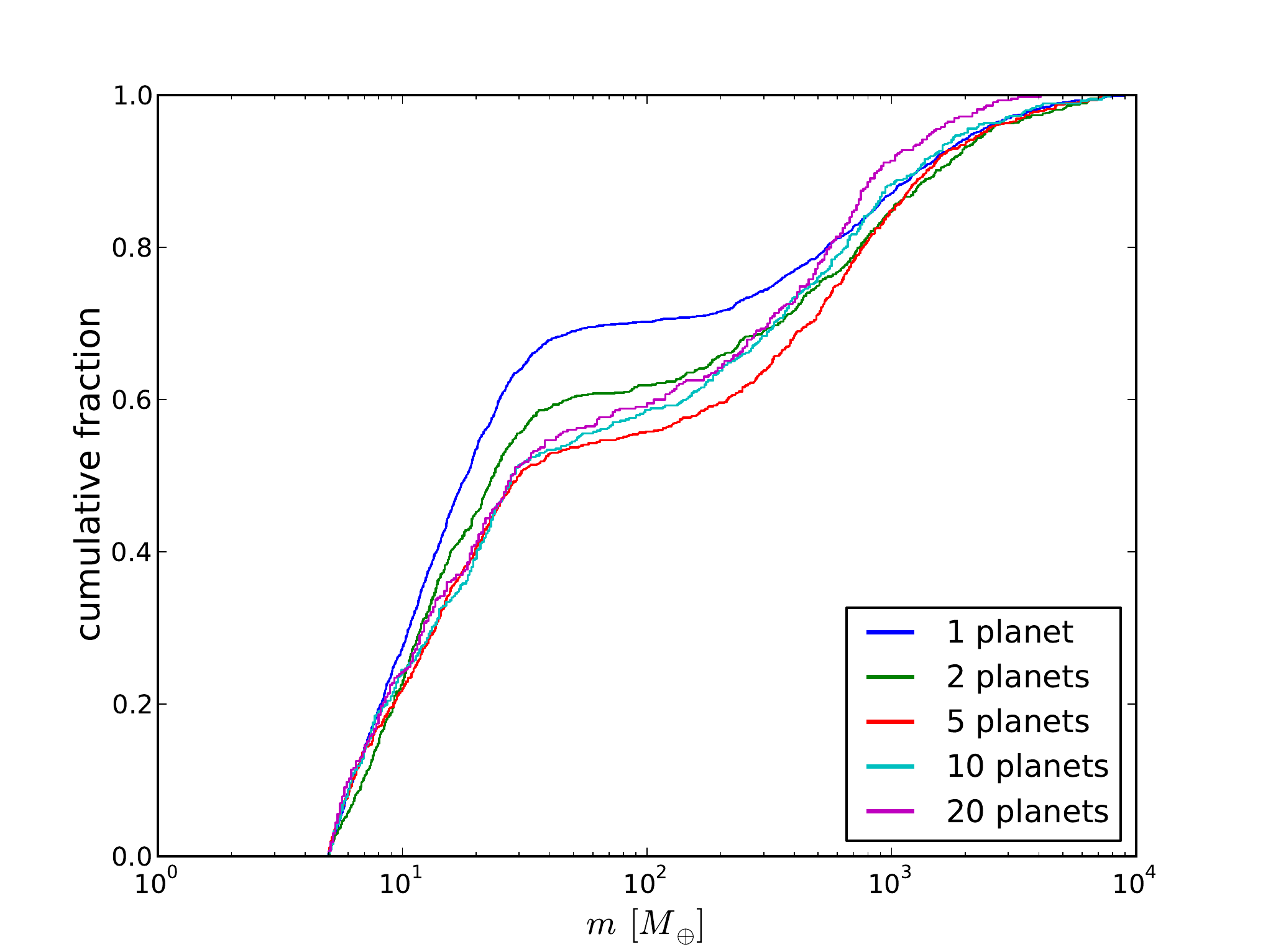}
    \includegraphics[width=0.3\textheight]{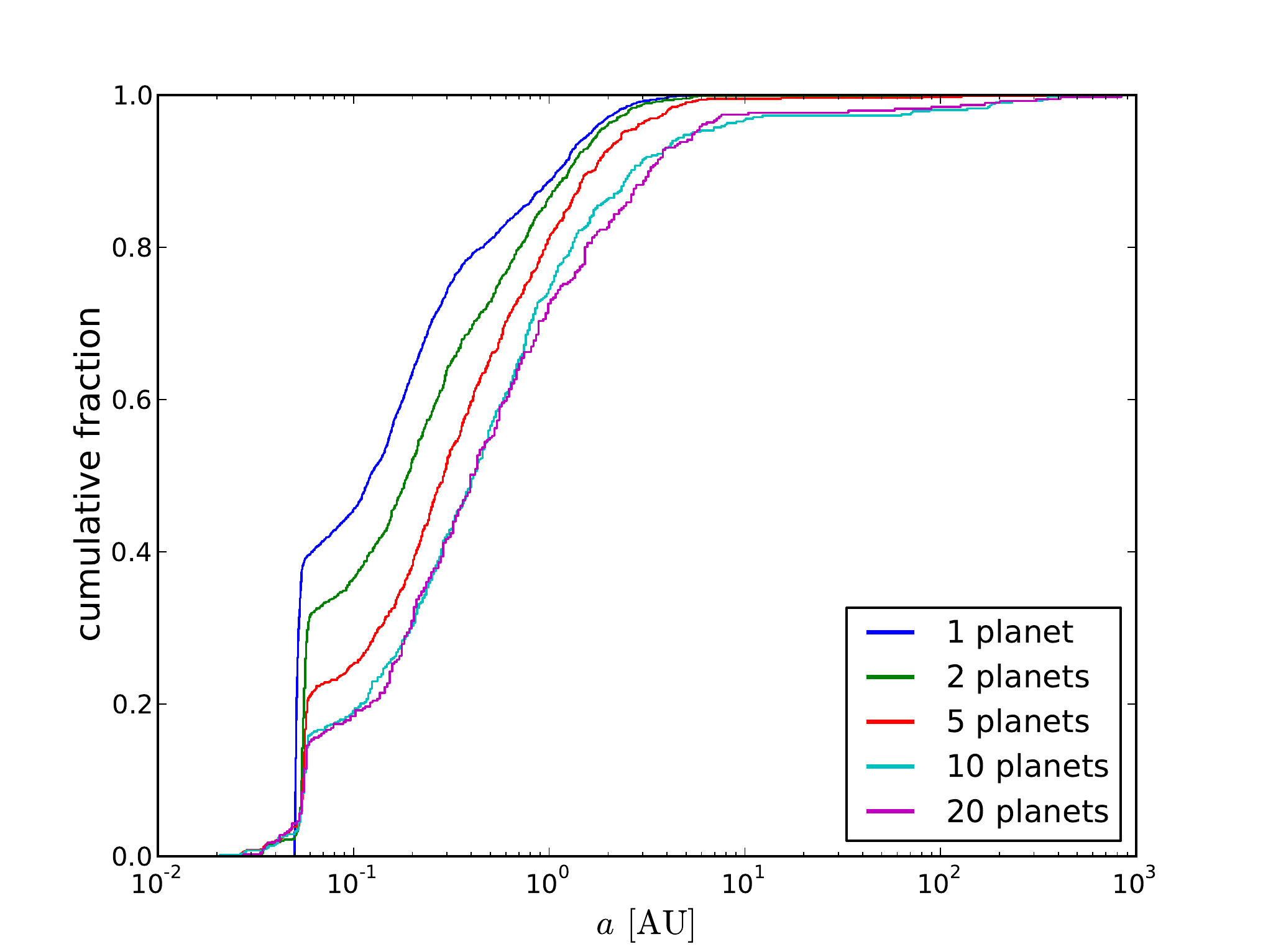}

  \caption{Top: Mean number of planets per system more massive than a given value, for different simulations, assuming different numbers of planetary embryos
  initially present in the system.  Middle: cumulative mass function, considering only planets more massive than 
  $5 \mearth$ still present in the system at the end of the simulation (planets colliding with the central
  star or ejected are not considered). Bottom: cumulative distribution of
  semi-major axis for the same population as in the middle panel. The number of planetary embryos assumed in each set of simulation is indicated on the panels. 
  Note, on the bottom panel, the planets that have been transported
  inside 0.05 AU by gravitational interactions with other planets of the same system.}
  \label{nbplanets}
\end{figure}

In order to quantify the effect of the initial number of planetary embryos, we have also computed the average number (per system) of planets larger than a given value,
and compared the results for the same set of simulations (assuming 1, 2, 5, 10, or 20 planetary embryos are initially present in the same disc). As before, the initial
locations of planetary embryos are statistically similar in all the cases. 
As can be seen in Fig.~\ref{nbplanets} (top panel), all the curves converge for large mass planets, but diverge towards the lower mass end. 
Simulations assuming a larger number of planetary embryos tend to lead to the formation of more low mass planets, which is somewhat expected. 
Another interesting aspect is that one notes a convergence of the curves as the number of planetary embryos increases. We can conclude for example that simulations
assuming 10 or 20 planetary embryos lead to similar results if one considers only planets more massive than a few $\mearth$. 

Considering the cumulative distribution of planetary masses (Fig.~\ref{nbplanets}, middle panel), taking into account objects more massive than 5 $\mearth$, it is clear that the mass function converges for more than 10 planetary
embryos (see the light blue and purple lines). The distribution of semi-major axis presents also a convergence for more than 10 planetary embryos (Fig. \ref{nbplanets}, bottom panel).
Note that only planets that are present in the system at the end of the simulation are considered in this plot. Planets ejected or accreted by the central star are not taken into
account in the calculation. If one considers now all the planets, the fraction of ejected planets increases monotonically with the number of planetary embryos initially
present in the simulation, from nearly 0 for 2 planetary embryos, to 3\%, 5\%, and 8\% for 5, 10, and 20 planetary embryos respectively.


\begin{figure}
  \center

  \includegraphics[width=0.35\textheight]{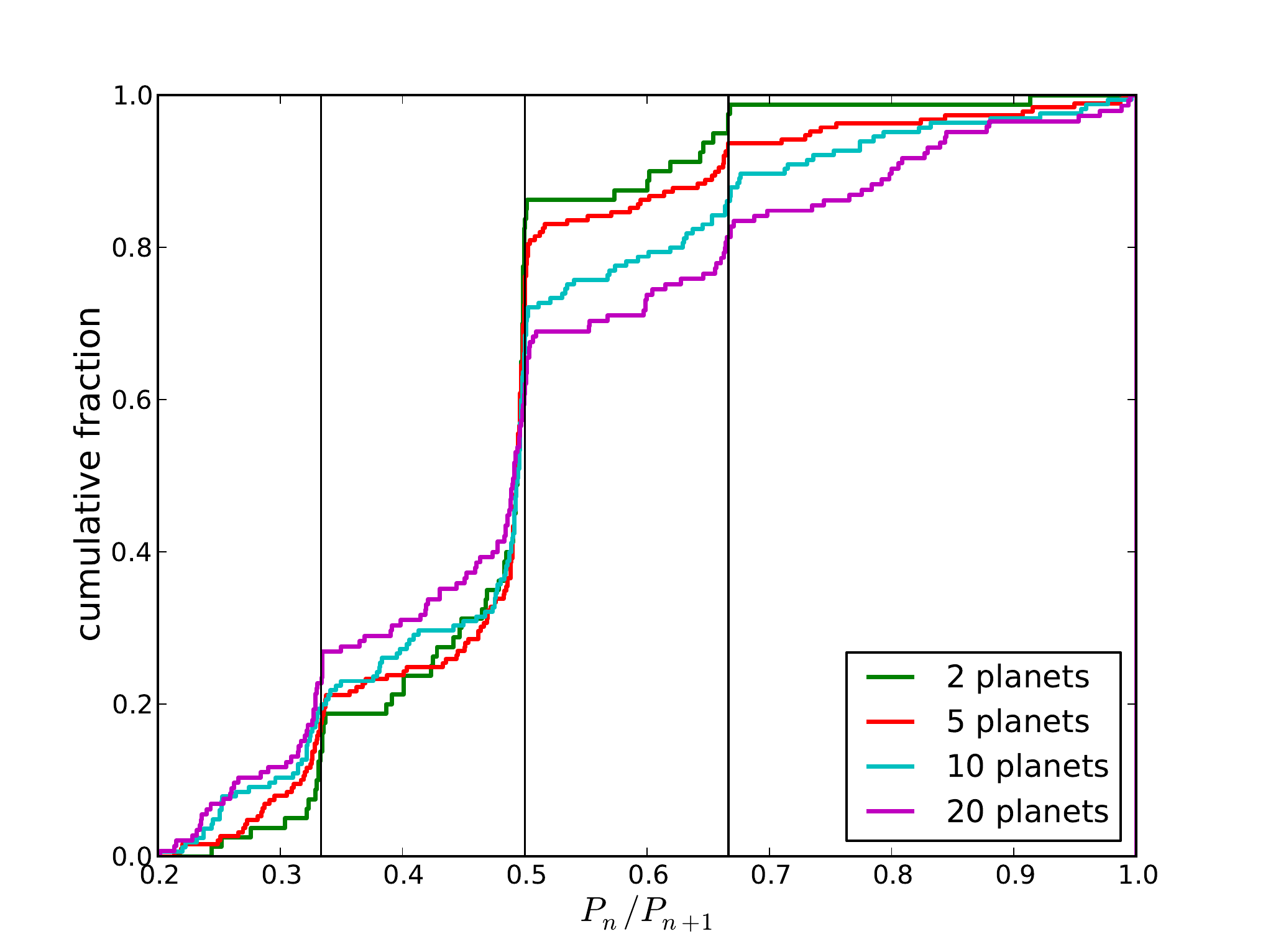}
    
  \caption{Period ratio of all pairs of planets more massive than 5 $\mearth$, for different simulations assuming initially
  different numbers of planetary embryos. The vertical lines show the location of the most important mean motion resonances.}
  \label{period_ratio}
\end{figure}

It is also interesting to compare the period ratios we obtain, as a function of the number of planetary embryos initially present in the system. Fig. \ref{period_ratio}
presents the period ratios of all planets more massive than $ 5 \mearth$, for the different simulations presented above (starting
with 5, 10, or 20 planetary embryos). As can be seen on the figure, the importance of the mean motion resonances decreases when the number of planetary
embryos increases~\citep[see also][]{2012MNRAS.427L..21R}: in the case of 2 planetary embryos, nearly all the systems end in mean motion resonance, whereas this fraction is
much smaller in the case of 20 planetary embryos. Contrary to the mass and semi-major cumulative histograms, there is still a difference between the cases with
10 and 20 planetary embryos, which means that the precise architecture of planetary systems depends on the amount of planetary embryos assumed to be present
in the system. 


%

\subsection{Eccentricity and inclination damping}

\begin{figure}
  \center

  \includegraphics[width=0.25\textheight,angle=-90]{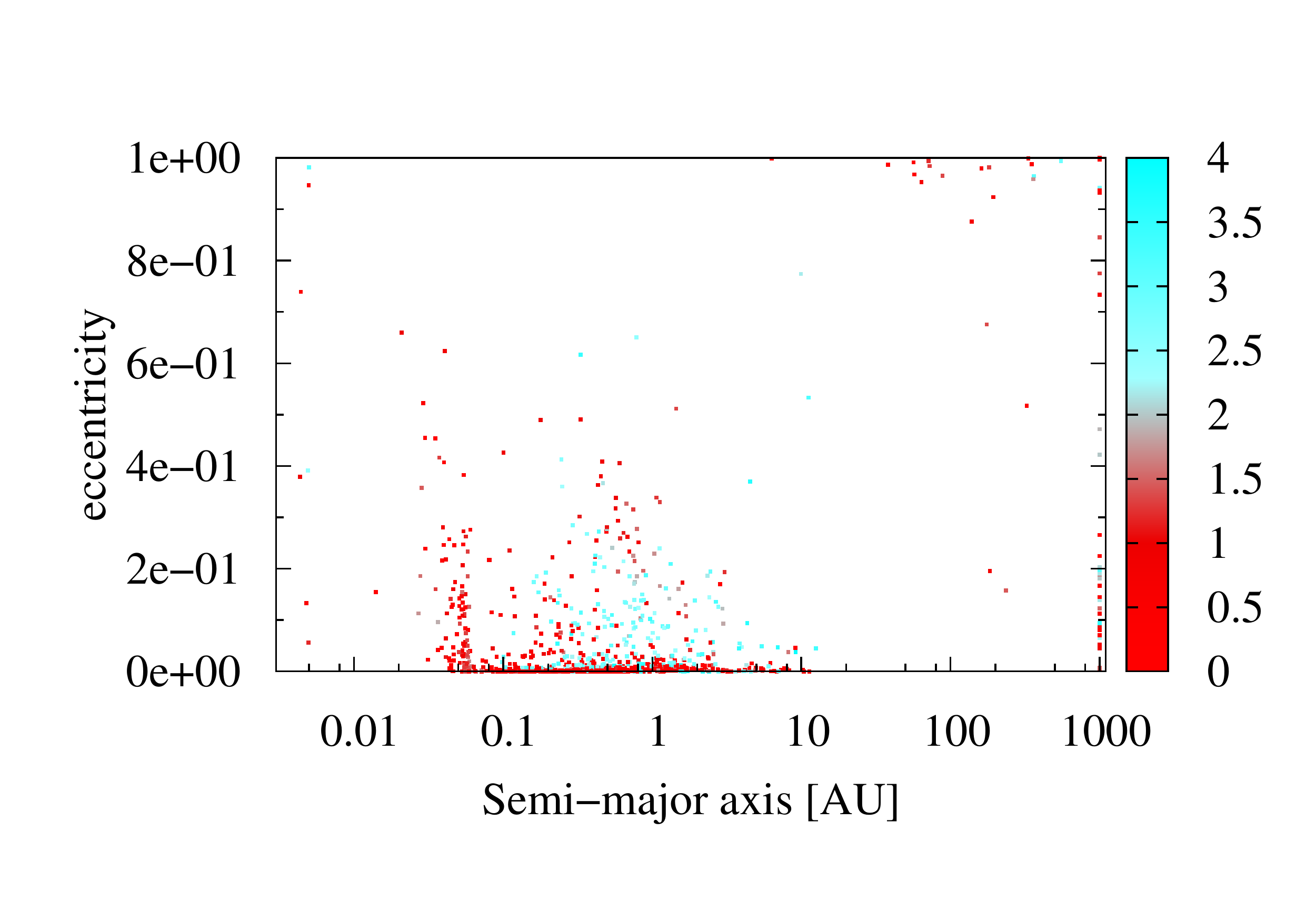}
  \includegraphics[width=0.25\textheight,angle=-90]{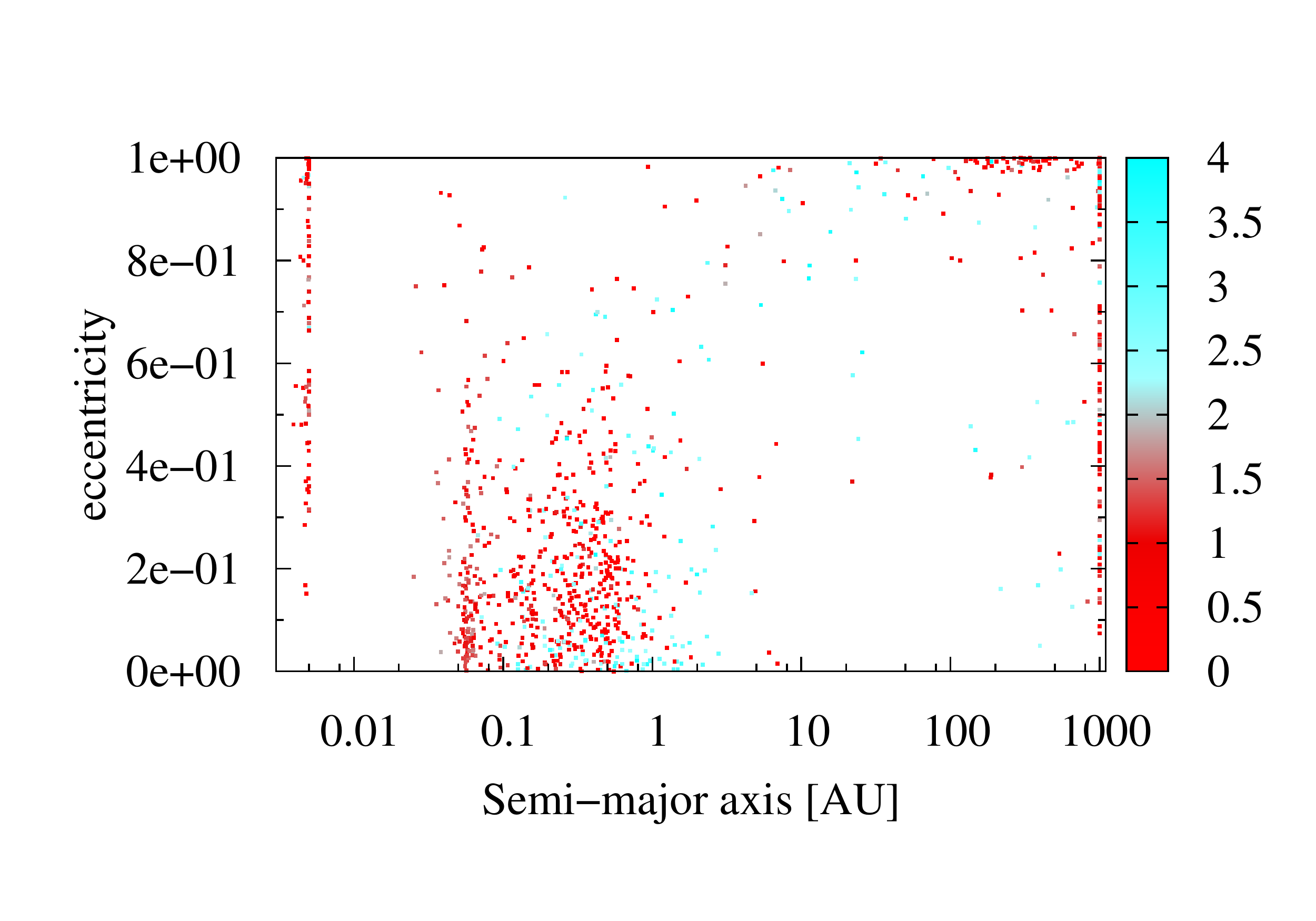}
  \includegraphics[width=0.25\textheight,angle=-90]{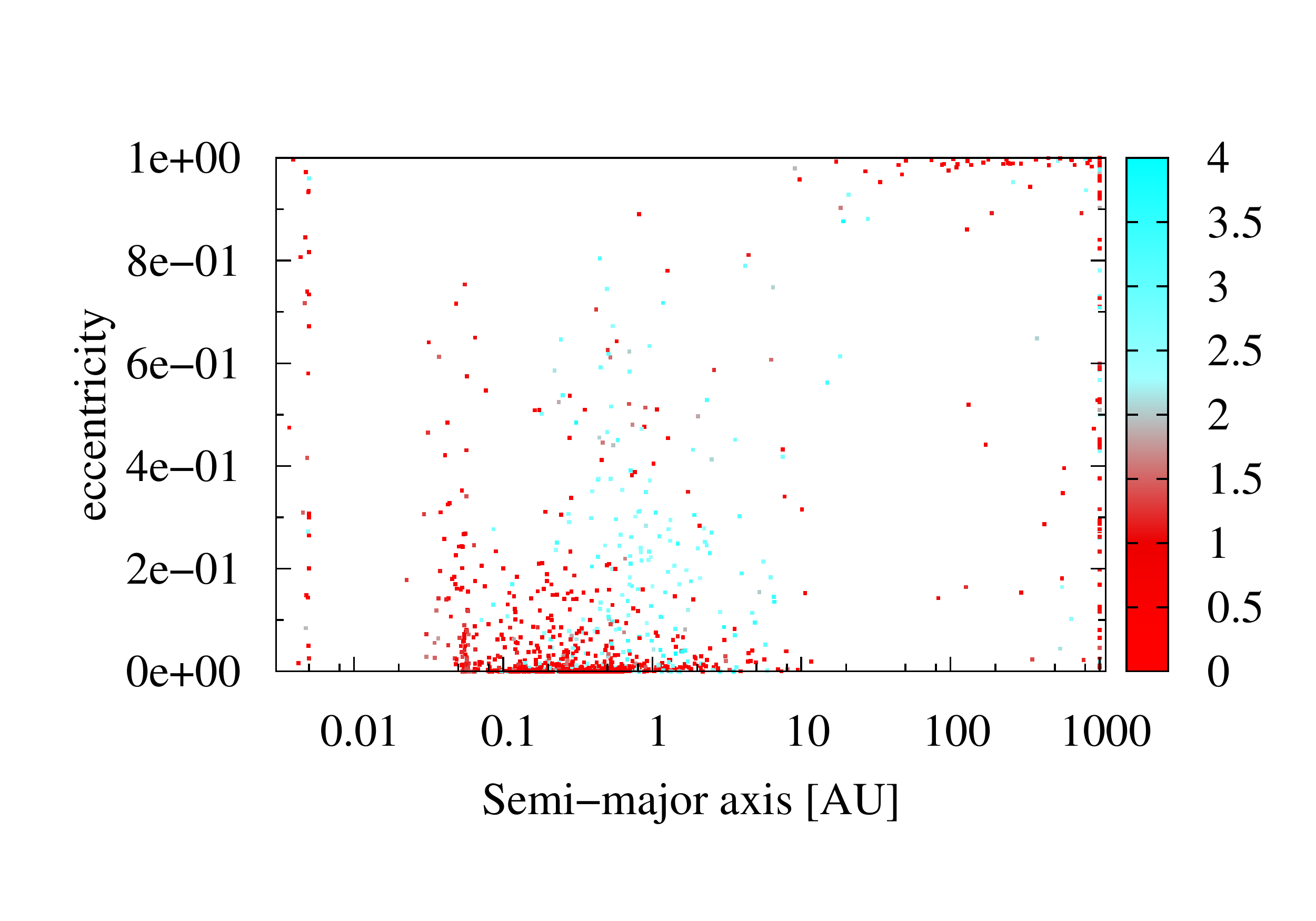}

  \caption{Semi-major axis \textit{versus} eccentricity for three models. Top: nominal 10-planets model, middle: without eccentricity and inclination damping, bottom: with eccentricity
  and inclination damping timescales increased by a factor 10 compared to the nominal model. The color code indicates the mass of the planet, in Earth masses (in log scale). }
  \label{ae_CD2133}
\end{figure}

\begin{figure}
  \center

  \includegraphics[width=0.25\textheight,angle=-90]{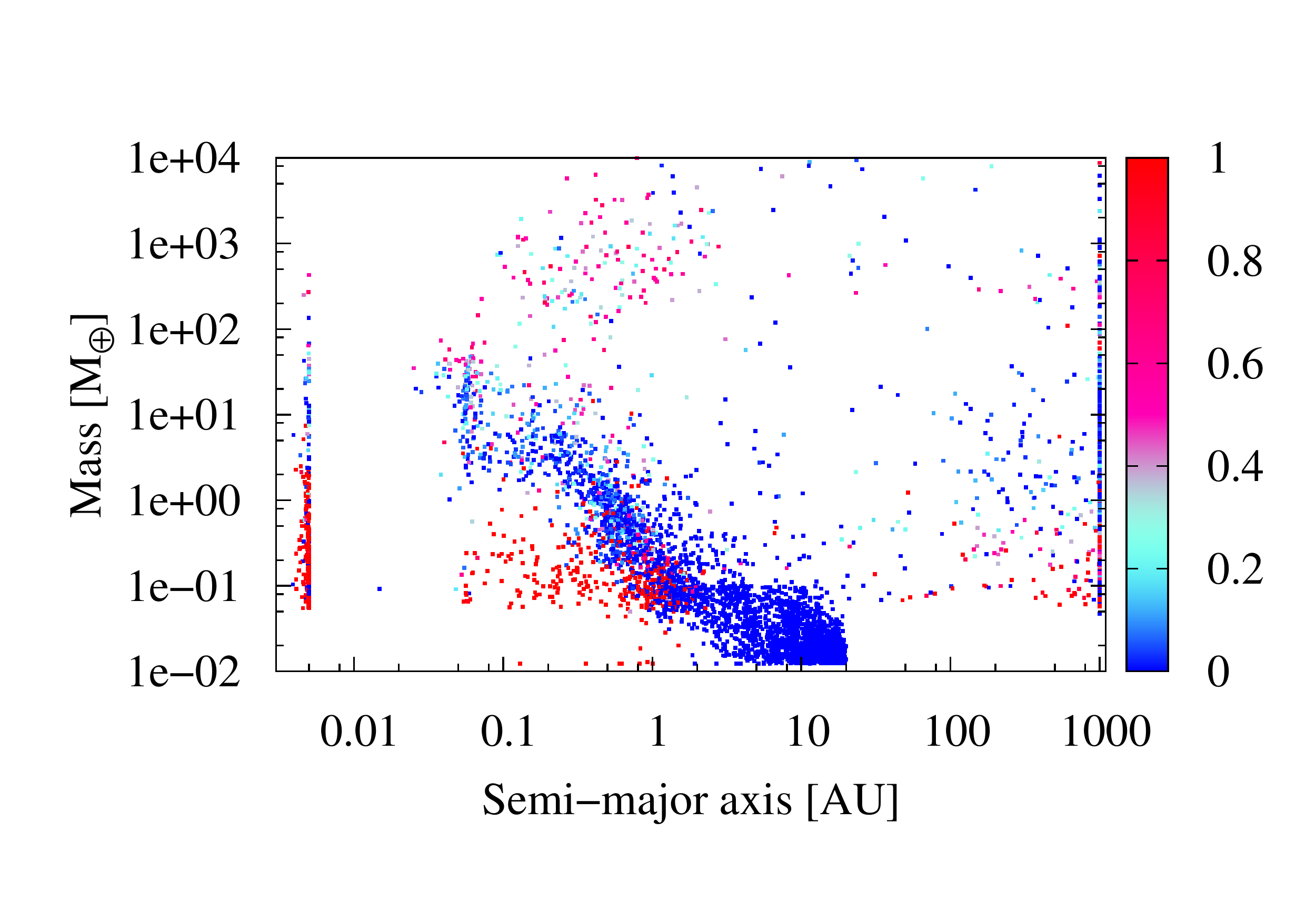}
  \includegraphics[width=0.25\textheight,angle=-90]{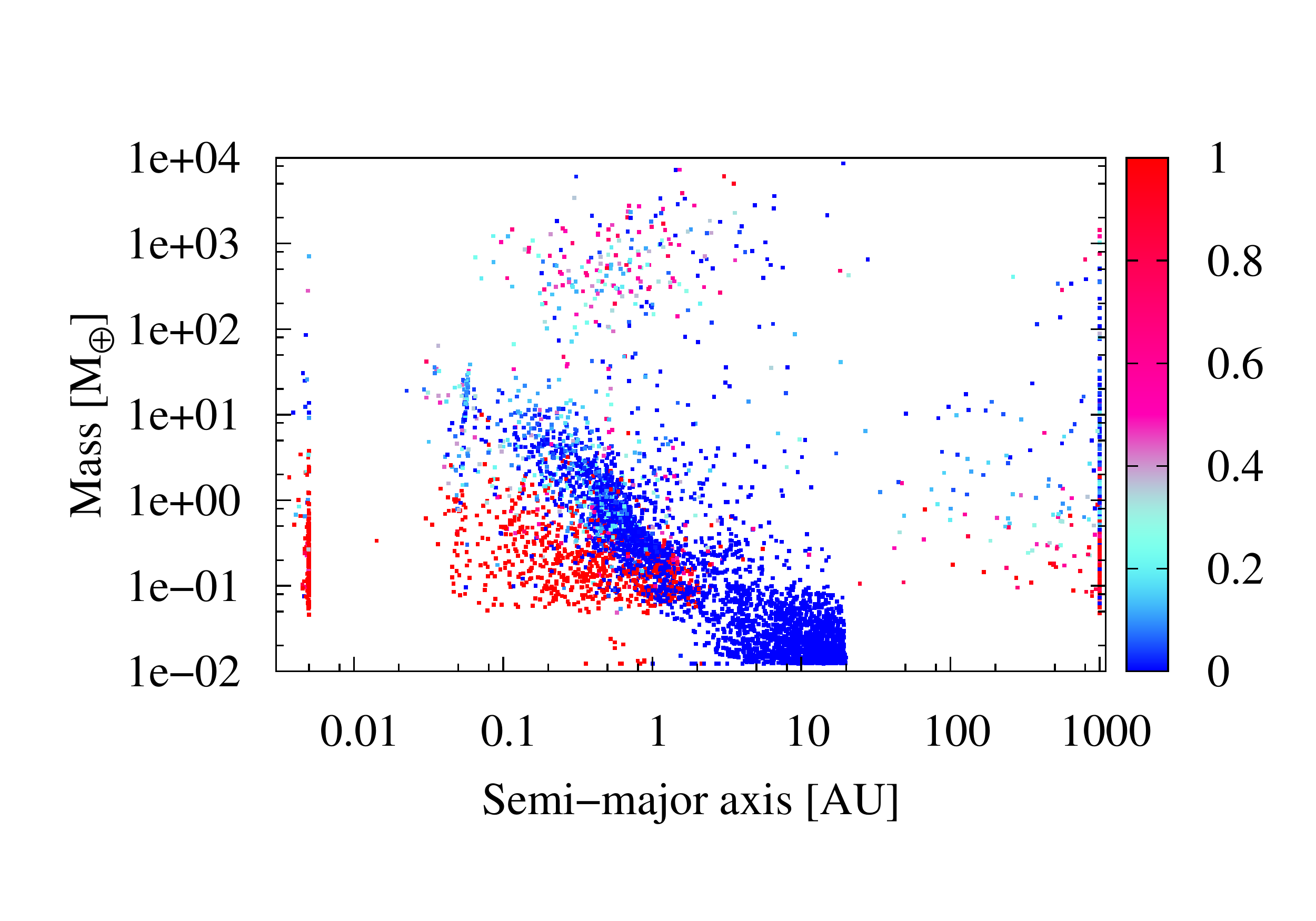}
  \caption{Same model as in Fig. \ref{am_systeme}, with modified eccentricity and inclination damping for planets. Top: no damping, bottom: damping timescale increased
  by 10 compared to the nominal case. The number of points are respectively 4650 and 5030. }
  \label{am_CD2156}
\end{figure}

We have also tested the effect of the timescale of planet's eccentricity and inclination damping, assuming a damping timescale
increased by a factor 10, or no damping at all (Fig. \ref{am_CD2156}).
This simulation resulted obviously in an increased mean eccentricity of planets, at the end of the formation process (see Fig. \ref{ae_CD2133}).
However, the masses and semi-major axis of planets that survived the formation where not too different from the standard 10-planet
case presented in Fig. \ref{am_systeme}. The main difference can be seen in the population of planets at distances between 10 and 100
AUs, and the number of ejected planets, which are more numerous in the low damping case. Interestingly enough, the comparison between
the number of planets at these distances, with the results of future direct imaging surveys, could put some constraints on these
components of the model.


{Interestingly enough, the period ratios obtained with less damping, or without any damping
of eccentricity and inclination seem to be closer to the ones observed by Kepler (in particular, the fraction of planets close to mean-motion resonance
is decreased when the damping is reduced). Indeed, in our nominal 10-planet case a larger fraction of planets than observed find themselves at mean motion resonances at the end of their formation. This suggests that the eccentricity and inclination damping is overestimated in our models. A more detailed analysis of this effect, as well as comparisons with Kepler results, will be presented in a forthcoming paper.}

\subsection{Limitations of the model}

As all theoretical models, the one presented in this paper is limited in certain aspects. In order to put our results in perspective,
we list now some of the most important assumptions and limitations. This will also provide a list of work we intend
to develop in the future.

\subsubsection{Planetesimal disc}

The numerical treatment of the disc of planetesimals is, in this work, simple, as the characteristics of planetesimals
only depend on their semi-major axis. This is the case for the mass (or radius), as well as for the eccentricity and inclination.
More specifically, we compute the evolution of the r.m.s eccentricity and inclination of planetesimals, assuming that they are
well described by a Rayleigh distribution.

This approach, although it constitutes an improvement with regard to former models~\citepalias[e.g.][]{2005A&A...434..343A,2009A&A...501.1139M,2009A&A...501.1161M,2012A&A...547A.111M}, where the excitation
and damping  of planetesimals by forming planets and gas drag was not computed accurately, is limited in the sense that some important
processes are not included. Among them, one can cite the orbital drifting of planetesimals, due to gas drag, as well as the formation
of a gap in the planetesimal disc. 

Moreover, as already mentioned, the mass of planetesimals at a given radius does not evolve with time, implying that fragmentation
and mass growth of planetesimals are not included. We plan to improve these aspects by using a model similar to the
one recently proposed by~\citet{2012ApJ...747..115O}. Finally, planetesimals have in this model no effect on the orbital evolution of planets: the planetesimal driven migration~\citep[e.g.][]{2012ApJ...758...80O} and the damping of eccentricity and inclination by planetesimals are not included in the model. These effects could indeed be potentially very important in regions of the disc where the gas surface density is small (e.g. outer parts of the disc), or at the end of the disc lifetime, and during long term evolution. 

\subsubsection{Planet-planet interaction through the protoplanetary disc}

As mentioned at the beginning of this paper, some interactions between planets are mediated by the
gas phase of the protoplanetary disc. Indeed, a first planet, if massive enough, is likely to modify in a substantial
way the gas surface density, therefore modifying angular momentum exchange and migration. Such effects
have been studied in different papers \citep[e.g.][]{2001MNRAS.320L..55M,2007Icar..191..158M}.
In particular, it is suspected that, when two planets grow massive enough to open a gap in the disc, and when
these two gaps merge, the angular momentum evolution of the global system can lead, under certain conditions
(e.g. related to the mass ratio between the two planets), to outward migration of both planets, catching them in
mean motion resonance. This process is in particular believed to have been at work during the late stage
of the formation of the Solar System~\citep[e.g.][]{2011Natur.475..206W}. We plan to investigate specifically the formation 
of the Solar System in a forthcoming paper.

\subsubsection{Long term evolution}

As already mentioned above, we focus in this formation model on the mass growth and orbital evolution of
planets in planetary systems during the existence of the gas phase of the protoplanetary disc. The reason for this limitation is that we
consider mainly the formation of planets with a non negligible gas envelope, whose growth is stopped when
the gas disc has disappeared. Moreover, a large fraction of planetesimals have
been either ejected or accreted by planets at the end of the protoplanetary (gas) disc life, when we stop the computation.
We therefore expect that planets should not notably grow after this period, except as a result of collisions between planets
or planetary embryos.

The {disappearance} of the protoplanetary disc has however not only the consequence of stopping the mass growth (in
term of gas) of planets, but it also means the end of eccentricity and inclination damping of planets. As a result, the
dynamical state of planetary systems is likely to evolve, leading to a re-arrangement of the global architecture of
systems. We have performed test calculations of the evolution of planetary systems presented in this paper, and have
found that the mass and semi-major axis of planets are not strongly modified, at least at the population level.
The long-term interaction between planets, however, has an effect in the increase of planetary eccentricities. Such
calculations will be presented in a forthcoming paper~(Pfyffer~et~al.,~in~prep).

\subsubsection{Planetary internal structure}

The models presented here are derived closely from the former models presented in~\citetalias{2005A&A...434..343A,2012A&A...547A.112M,2012A&A...547A.111M}~and~\citetalias{2013A&A...549A..44F}. In particular, it is assumed, when we compute
the internal structure of forming planets, that all accreted planetesimals reach the planetary core, {depositing} at this location their mass and energy.
It is however likely that planetesimals, in particular those of low mass, are destroyed during their travel towards the planetary center. This
leads to a modification of the core luminosity, a reduced core growth, and an increase of the metallicity of the planetary envelope.

It has been recently shown~\citep{2011MNRAS.416.1419H}, that a change in the mean opacity and equation of state in the planetary envelope, itself resulting from the destruction of incoming planetesimals, can  heavily modify the planetary critical mass, and therefore the whole
planet formation timescale. The afore-mentioned study, however, assumed some value of the metallicity in the planetary envelope,
which is not computed as a result of planet formation. It is however more likely that the metallicity of planets will change with time,
as a result of the accretion of planetesimals, and as a function of the stability of the envelope with regards to convection. Indeed,
if convection is efficient enough, the heavy elements deposited by planetesimals are likely to be equally distributed in the whole convective
zone, whereas heavy elements and grains could settle down in the radiative zone. Such a self-consistent computation of the planetary
internal structure and its effect on the planetary growth and migration will be studied in a forthcoming paper.

\subsection{Conclusion}

{We have extended our planet formation model to include the formation of planetary systems. For this, we seed our simulations with a number (ranging from 2 to 20) of small seed embryos. We show that the presence of several growing embryos can result in very important modifications in the overall formation process. In particular, gravitational interactions between these growing bodies results in significant changes in the final mass, semi-major axis and orbital parameters, in particular as a result of the larger orbital migration of planets, which itself results from planet-planet interactions. As a result, planets belonging to planetary systems are found to be more water rich in the region around 1 AU, and a population of low mass, close-in planets, which is not present when considering the growth of only one planet, appears.}

{We have also demonstrated that the mass distribution and cumulative distribution of planets do not strongly depend on the number of planetary embryos considered, in particular for planets more massive than a few Earth masses. However, the distribution of period-ratios between planets does depend on the number of planetary embryos, even if the dependance seems to decrease with the number of embryos. The distribution of period ratios also shows that our model predicts too many systems in, or close to, mean-motion resonance. This could result from effects that are not taken into account in our model, for example from stochastic effects~\citep[see e.g.][]{2012MNRAS.427L..21R}, or from an overestimation, in our model, of the eccentricity and inclination damping. Future work will address these issues, as well as the ones presented in the previous sections.}

\acknowledgements

This work was supported by the European Research Council under grant 239605, the Swiss National Science Foundation, and the Alexander von Humboldt Foundation.

\bibliographystyle{aa}
\bibliography{lit}

\begin{thebibliography}{45}
\expandafter\ifx\csname natexlab\endcsname\relax\def\natexlab#1{#1}\fi

\bibitem[{{Alibert} {et~al.}(2006){Alibert}, {Baraffe}, {Benz}, {Chabrier},
  {Mordasini}, {Lovis}, {Mayor}, {Pepe}, {Bouchy}, {Queloz}, \&
  {Udry}}]{2006A&A...455L..25A}
{Alibert}, Y., {Baraffe}, I., {Benz}, W., {et~al.} 2006, \aap, 455, L25

\bibitem[{{Alibert} {et~al.}(2005{\natexlab{a}}){Alibert}, {Mordasini}, {Benz},
  \& {Winisdoerffer}}]{2005A&A...434..343A}
{Alibert}, Y., {Mordasini}, C., {Benz}, W., \& {Winisdoerffer}, C.
  2005{\natexlab{a}}, \aap, 434, 343

\bibitem[{{Alibert} {et~al.}(2005{\natexlab{b}}){Alibert}, {Mousis},
  {Mordasini}, \& {Benz}}]{2005ApJ...626L..57A}
{Alibert}, Y., {Mousis}, O., {Mordasini}, C., \& {Benz}, W. 2005{\natexlab{b}},
  \apjl, 626, L57

\bibitem[{{Andrews} {et~al.}(2010){Andrews}, {Wilner}, {Hughes}, {Qi}, \&
  {Dullemond}}]{2010ApJ...723.1241A}
{Andrews}, S.~M., {Wilner}, D.~J., {Hughes}, A.~M., {Qi}, C., \& {Dullemond},
  C.~P. 2010, \apj, 723, 1241

\bibitem[{{Borucki} {et~al.}(2011){Borucki}, {Koch}, {Basri}, {Batalha},
  {Boss}, {Brown}, {Caldwell}, {Christensen-Dalsgaard}, {Cochran}, {DeVore},
  {Dunham}, {Dupree}, {Gautier}, {Geary}, {Gilliland}, {Gould}, {Howell},
  {Jenkins}, {Kjeldsen}, {Latham}, {Lissauer}, {Marcy}, {Monet}, {Sasselov},
  {Tarter}, {Charbonneau}, {Doyle}, {Ford}, {Fortney}, {Holman}, {Seager},
  {Steffen}, {Welsh}, {Allen}, {Bryson}, {Buchhave}, {Chandrasekaran},
  {Christiansen}, {Ciardi}, {Clarke}, {Dotson}, {Endl}, {Fischer}, {Fressin},
  {Haas}, {Horch}, {Howard}, {Isaacson}, {Kolodziejczak}, {Li}, {MacQueen},
  {Meibom}, {Prsa}, {Quintana}, {Rowe}, {Sherry}, {Tenenbaum}, {Torres},
  {Twicken}, {Van Cleve}, {Walkowicz}, \& {Wu}}]{2011ApJ...728..117B}
{Borucki}, W.~J., {Koch}, D.~G., {Basri}, G., {et~al.} 2011, \apj, 728, 117

\bibitem[{{Broeg} \& {Benz}(2012)}]{2012A&A...538A..90B}
{Broeg}, C.~H. \& {Benz}, W. 2012, \aap, 538, A90

\bibitem[{{Carron}(2013)}]{2013PhD......Carron}
{Carron}, F. 2013, PhD thesis, University of Bern

\bibitem[{{Crida} {et~al.}(2006){Crida}, {Morbidelli}, \&
  {Masset}}]{2006Icar..181..587C}
{Crida}, A., {Morbidelli}, A., \& {Masset}, F. 2006, \icarus, 181, 587

\bibitem[{{Fogg} \& {Nelson}(2007)}]{2007A&A...472.1003F}
{Fogg}, M.~J. \& {Nelson}, R.~P. 2007, \aap, 472, 1003

\bibitem[{{Fortier} {et~al.}(2013){Fortier}, {Alibert}, {Carron}, {Benz}, \&
  {Dittkrist}}]{2013A&A...549A..44F}
{Fortier}, A., {Alibert}, Y., {Carron}, F., {Benz}, W., \& {Dittkrist}, K.-M.
  2013, \aap, 549, A44

\bibitem[{{Guilera} {et~al.}(2010){Guilera}, {Brunini}, \&
  {Benvenuto}}]{2010A&A...521A..50G}
{Guilera}, O.~M., {Brunini}, A., \& {Benvenuto}, O.~G. 2010, \aap, 521, A50

\bibitem[{{Guilera} {et~al.}(2011){Guilera}, {Fortier}, {Brunini}, \&
  {Benvenuto}}]{2011A&A...532A.142G}
{Guilera}, O.~M., {Fortier}, A., {Brunini}, A., \& {Benvenuto}, O.~G. 2011,
  \aap, 532, A142

\bibitem[{{Hori} \& {Ikoma}(2011)}]{2011MNRAS.416.1419H}
{Hori}, Y. \& {Ikoma}, M. 2011, \mnras, 416, 1419

\bibitem[{{Inaba} \& {Ikoma}(2003)}]{2003A&A...410..711I}
{Inaba}, S. \& {Ikoma}, M. 2003, \aap, 410, 711

\bibitem[{{Johansen} {et~al.}(2007){Johansen}, {Oishi}, {Mac Low}, {Klahr},
  {Henning}, \& {Youdin}}]{2007Natur.448.1022J}
{Johansen}, A., {Oishi}, J.~S., {Mac Low}, M.-M., {et~al.} 2007, \nat, 448,
  1022

\bibitem[{{Kalas} {et~al.}(2008){Kalas}, {Graham}, {Chiang}, {Fitzgerald},
  {Clampin}, {Kite}, {Stapelfeldt}, {Marois}, \& {Krist}}]{2008Sci...322.1345K}
{Kalas}, P., {Graham}, J.~R., {Chiang}, E., {et~al.} 2008, Science, 322, 1345

\bibitem[{{Lagrange} {et~al.}(2009){Lagrange}, {Gratadour}, {Chauvin}, {Fusco},
  {Ehrenreich}, {Mouillet}, {Rousset}, {Rouan}, {Allard}, {Gendron}, {Charton},
  {Mugnier}, {Rabou}, {Montri}, \& {Lacombe}}]{2009A&A...493L..21L}
{Lagrange}, A.-M., {Gratadour}, D., {Chauvin}, G., {et~al.} 2009, \aap, 493,
  L21

\bibitem[{{Lissauer} {et~al.}(2011){Lissauer}, {Fabrycky}, {Ford}, {Borucki},
  {Fressin}, {Marcy}, {Orosz}, {Rowe}, {Torres}, {Welsh}, {Batalha}, {Bryson},
  {Buchhave}, {Caldwell}, {Carter}, {Charbonneau}, {Christiansen}, {Cochran},
  {Desert}, {Dunham}, {Fanelli}, {Fortney}, {Gautier}, {Geary}, {Gilliland},
  {Haas}, {Hall}, {Holman}, {Koch}, {Latham}, {Lopez}, {McCauliff}, {Miller},
  {Morehead}, {Quintana}, {Ragozzine}, {Sasselov}, {Short}, \&
  {Steffen}}]{2011Natur.470...53L}
{Lissauer}, J.~J., {Fabrycky}, D.~C., {Ford}, E.~B., {et~al.} 2011, \nat, 470,
  53

\bibitem[{{Lovis} {et~al.}(2011){Lovis}, {S{\'e}gransan}, {Mayor}, {Udry},
  {Benz}, {Bertaux}, {Bouchy}, {Correia}, {Laskar}, {Lo Curto}, {Mordasini},
  {Pepe}, {Queloz}, \& {Santos}}]{2011A&A...528A.112L}
{Lovis}, C., {S{\'e}gransan}, D., {Mayor}, M., {et~al.} 2011, \aap, 528, A112

\bibitem[{{Mamajek}(2009)}]{2009AIPC.1158....3M}
{Mamajek}, E.~E. 2009, in American Institute of Physics Conference Series, Vol.
  1158, American Institute of Physics Conference Series, ed. T.~{Usuda},
  M.~{Tamura}, \& M.~{Ishii}, 3--10

\bibitem[{{Marois} {et~al.}(2008){Marois}, {Macintosh}, {Barman}, {Zuckerman},
  {Song}, {Patience}, {Lafreni{\`e}re}, \& {Doyon}}]{2008Sci...322.1348M}
{Marois}, C., {Macintosh}, B., {Barman}, T., {et~al.} 2008, Science, 322, 1348

\bibitem[{{Masset} \& {Snellgrove}(2001)}]{2001MNRAS.320L..55M}
{Masset}, F. \& {Snellgrove}, M. 2001, \mnras, 320, L55

\bibitem[{{Masset} {et~al.}(2006){Masset}, {Morbidelli}, {Crida}, \&
  {Ferreira}}]{2006ApJ...642..478M}
{Masset}, F.~S., {Morbidelli}, A., {Crida}, A., \& {Ferreira}, J. 2006, \apj,
  642, 478

\bibitem[{{Mayor} {et~al.}(2011){Mayor}, {Marmier}, {Lovis}, {Udry},
  {S{\'e}gransan}, {Pepe}, {Benz}, {Bertaux}, {Bouchy}, {Dumusque}, {Lo Curto},
  {Mordasini}, {Queloz}, \& {Santos}}]{2011arXiv1109.2497M}
{Mayor}, M., {Marmier}, M., {Lovis}, C., {et~al.} 2011, {\it astroph} 1109.2497

\bibitem[{{Mayor} \& {Queloz}(1995)}]{1995Natur.378..355M}
{Mayor}, M. \& {Queloz}, D. 1995, \nat, 378, 355

\bibitem[{{Min} {et~al.}(2011){Min}, {Dullemond}, {Kama}, \&
  {Dominik}}]{2011Icar..212..416M}
{Min}, M., {Dullemond}, C.~P., {Kama}, M., \& {Dominik}, C. 2011, \icarus, 212,
  416

\bibitem[{{Morbidelli} \& {Crida}(2007)}]{2007Icar..191..158M}
{Morbidelli}, A. \& {Crida}, A. 2007, \icarus, 191, 158

\bibitem[{{Mordasini} {et~al.}(2009{\natexlab{a}}){Mordasini}, {Alibert}, \&
  {Benz}}]{2009A&A...501.1139M}
{Mordasini}, C., {Alibert}, Y., \& {Benz}, W. 2009{\natexlab{a}}, \aap, 501,
  1139

\bibitem[{{Mordasini} {et~al.}(2009{\natexlab{b}}){Mordasini}, {Alibert},
  {Benz}, \& {Naef}}]{2009A&A...501.1161M}
{Mordasini}, C., {Alibert}, Y., {Benz}, W., \& {Naef}, D. 2009{\natexlab{b}},
  \aap, 501, 1161

\bibitem[{{Mordasini} {et~al.}(2012{\natexlab{a}}){Mordasini}, {Alibert},
  {Georgy}, {Dittkrist}, {Klahr}, \& {Henning}}]{2012A&A...547A.112M}
{Mordasini}, C., {Alibert}, Y., {Georgy}, C., {et~al.} 2012{\natexlab{a}},
  \aap, 547, A112

\bibitem[{{Mordasini} {et~al.}(2012{\natexlab{b}}){Mordasini}, {Alibert},
  {Klahr}, \& {Henning}}]{2012A&A...547A.111M}
{Mordasini}, C., {Alibert}, Y., {Klahr}, H., \& {Henning}, T.
  2012{\natexlab{b}}, \aap, 547, A111

\bibitem[{{Mordasini} {et~al.}(2011){Mordasini}, {Dittkrist}, {Alibert},
  {Klahr}, {Benz}, \& {Henning}}]{2011IAUS..276...72M}
{Mordasini}, C., {Dittkrist}, K.-M., {Alibert}, Y., {et~al.} 2011, in IAU
  Symposium, Vol. 276, IAU Symposium, ed. A.~{Sozzetti}, M.~G. {Lattanzi}, \&
  A.~P. {Boss}, 72--75

\bibitem[{{Movshovitz} {et~al.}(2010){Movshovitz}, {Bodenheimer}, {Podolak}, \&
  {Lissauer}}]{2010Icar..209..616M}
{Movshovitz}, N., {Bodenheimer}, P., {Podolak}, M., \& {Lissauer}, J.~J. 2010,
  \icarus, 209, 616

\bibitem[{{Ormel} {et~al.}(2012){Ormel}, {Ida}, \&
  {Tanaka}}]{2012ApJ...758...80O}
{Ormel}, C.~W., {Ida}, S., \& {Tanaka}, H. 2012, \apj, 758, 80

\bibitem[{{Ormel} \& {Kobayashi}(2012)}]{2012ApJ...747..115O}
{Ormel}, C.~W. \& {Kobayashi}, H. 2012, \apj, 747, 115

\bibitem[{{Paardekooper} {et~al.}(2010){Paardekooper}, {Baruteau}, {Crida}, \&
  {Kley}}]{2010MNRAS.401.1950P}
{Paardekooper}, S.-J., {Baruteau}, C., {Crida}, A., \& {Kley}, W. 2010, \mnras,
  401, 1950

\bibitem[{{Paardekooper} {et~al.}(2011){Paardekooper}, {Baruteau}, \&
  {Kley}}]{2011MNRAS.410..293P}
{Paardekooper}, S.-J., {Baruteau}, C., \& {Kley}, W. 2011, \mnras, 410, 293

\bibitem[{{Pollack} {et~al.}(1996){Pollack}, {Hubickyj}, {Bodenheimer},
  {Lissauer}, {Podolak}, \& {Greenzweig}}]{1996Icar..124...62P}
{Pollack}, J.~B., {Hubickyj}, O., {Bodenheimer}, P., {et~al.} 1996, \icarus,
  124, 62

\bibitem[{{Rein}(2012)}]{2012MNRAS.427L..21R}
{Rein}, H. 2012, \mnras, 427, L21

\bibitem[{{Richardson} {et~al.}(2000){Richardson}, {Quinn}, {Stadel}, \&
  {Lake}}]{2000Icar..143...45R}
{Richardson}, D.~C., {Quinn}, T., {Stadel}, J., \& {Lake}, G. 2000, \icarus,
  143, 45

\bibitem[{{Shakura} \& {Sunyaev}(1973)}]{1973A&A....24..337S}
{Shakura}, N.~I. \& {Sunyaev}, R.~A. 1973, \aap, 24, 337

\bibitem[{{Tanaka} {et~al.}(2002){Tanaka}, {Takeuchi}, \&
  {Ward}}]{2002ApJ...565.1257T}
{Tanaka}, H., {Takeuchi}, T., \& {Ward}, W.~R. 2002, \apj, 565, 1257

\bibitem[{{Veras} \& {Armitage}(2004)}]{2004MNRAS.347..613V}
{Veras}, D. \& {Armitage}, P.~J. 2004, \mnras, 347, 613

\bibitem[{{Walsh} {et~al.}(2011){Walsh}, {Morbidelli}, {Raymond}, {O'Brien}, \&
  {Mandell}}]{2011Natur.475..206W}
{Walsh}, K.~J., {Morbidelli}, A., {Raymond}, S.~N., {O'Brien}, D.~P., \&
  {Mandell}, A.~M. 2011, \nat, 475, 206

\bibitem[{{Ward}(1997)}]{1997ApJ...482L.211W}
{Ward}, W.~R. 1997, \apjl, 482, L211

\end{thebibliography}
\end{document}